\documentclass[acmtog]{acmart}
\acmSubmissionID{102}

\usepackage{booktabs} 
\usepackage{amsmath}
\usepackage{amsfonts}
\usepackage{amssymb}
\usepackage{mathtools}
\usepackage{float}
\usepackage[toc,page]{appendix}
\DeclarePairedDelimiter{\ceil}{\lceil}{\rceil}

\usepackage[ruled]{algorithm2e} 

\SetAlFnt{\small}
\SetAlCapFnt{\small}
\SetAlCapNameFnt{\small}
\SetAlCapHSkip{0pt}

\newcommand{\ours}{MeshODE}

\begin{document}
\title{\ours{}: A Robust and Scalable Framework for Mesh Deformation}

\author{Jingwei Huang}
\authornote{Both authors contributed equally to the paper}
\affiliation{%
  \institution{Stanford University}
}
\author{Chiyu ``Max'' Jiang}
\authornotemark[1]
\affiliation{%
	\institution{University of California, Berkeley}
}
\author{Baiqiang Leng}
\affiliation{%
	\institution{Tsinghua University}
}
\author{Bin Wang}
\affiliation{%
	\institution{Tsinghua University}
}
\author{Leonidas Guibas}
\affiliation{%
	\institution{Stanford University}
}


\begin{abstract}
We present \ours{}, a scalable and robust framework for pairwise CAD model deformation without prespecified correspondences. Given a pair of shapes, our framework provides a novel shape feature-preserving mapping function that continuously deforms one model to the other by minimizing fitting and rigidity losses based on the non-rigid iterative-closest-point (ICP) algorithm.
We address two challenges in this problem, namely the design of a powerful deformation function and obtaining a feature-preserving CAD deformation.
While traditional deformation directly optimizes for the coordinates of the mesh vertices or the vertices of a control cage, we introduce a deep bijective mapping that utilizes a flow model parameterized as a neural network. Our function has the capacity to handle complex deformations, produces deformations that are guaranteed free of self-intersections, and requires low rigidity constraining for geometry preservation, which leads to a better fitting quality compared with existing methods. It additionally enables continuous deformation between two arbitrary shapes without supervision for intermediate shapes.
Furthermore, we propose a robust preprocessing pipeline for raw CAD meshes using feature-aware subdivision and a uniform graph template representation to address artifacts in raw CAD models including self-intersections, irregular triangles, topologically disconnected components, non-manifold edges, and nonuniformly distributed vertices. This facilitates a fast deformation optimization process that preserves global and local details.

On top of the methodological contributions, we create an evaluation benchmark for unsupervised pairwise shape deformation and find that our deformation results significantly outperform the state-of-the-art deformation algorithms with respect to fitting error.
We show that our framework benefits various downstream applications including novel shape design and animation, scan-to-CAD fitting, and texture transfer. Our code is publicly available\footnote{\url{https://github.com/hjwdzh/MeshODE}}.
\end{abstract}

\begin{teaserfigure}
    \centering
    \includegraphics[width=\textwidth]{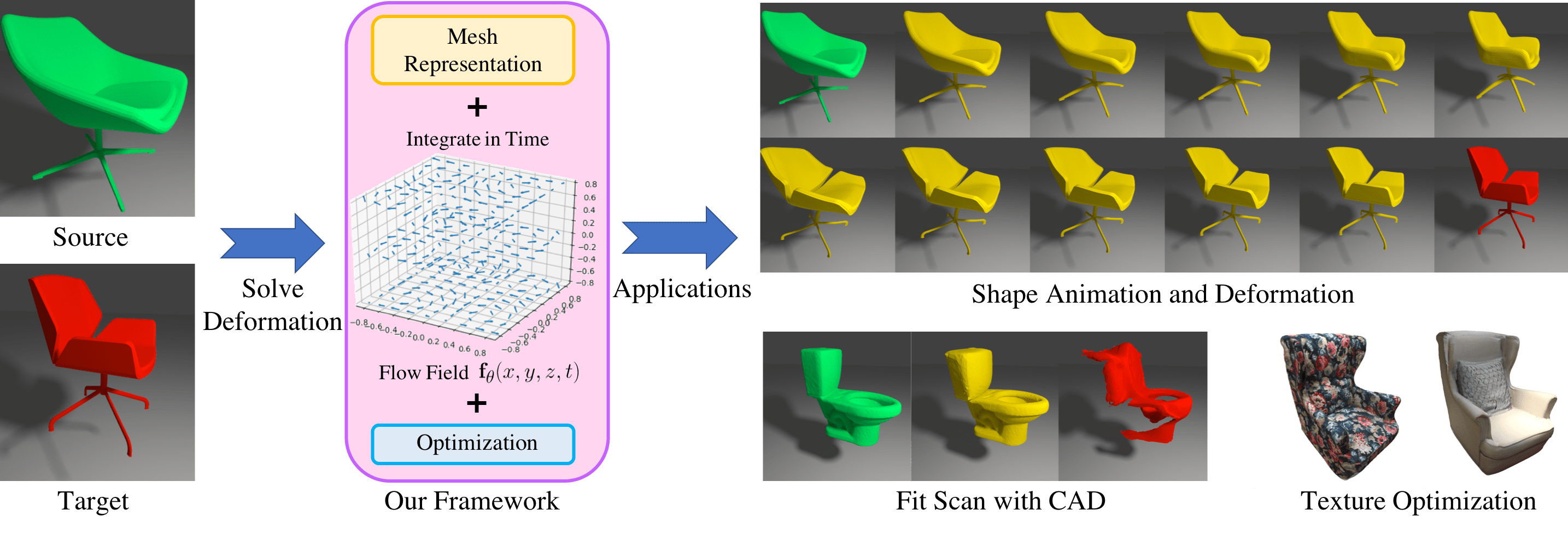}
    \caption{\ours{} is a robust and scalable framework that addresses the problem of pairwise shape deformation without prespecified correspondences. It enables a range of computer graphics and computer vision applications including shape animation and creation, scan-to-CAD fitting and texture reconstruction.}
    \label{fig:teaser}
\end{teaserfigure}

\maketitle

\section{Introduction}
\paragraph{The Shape Deformation Problem} 3D shape deformation is a fundamental problem in computer graphics and computer vision with various applications in these fields.
For starters, shape deformation as a means of geometry editing can be useful for creating novel digital content by deforming a preexisting model. \cite{sorkine2007rigid,jiang2017consistent} deform a source 3D model to a target shape given pairs of sparse correspondences between the source and the target to create novel shape designs. Such deformation can also be transferred to novel source shapes~\cite{sumner2004deformation,yifan2019neural}.
Moreover, shape deformation can be used for computer animation~\cite{anguelov2005scape,jacobson2011stretchable} by creating smooth interpolations between key-frames.
Recently, \cite{yifan2019neural,wang20193dn} addressed the more challenging problem of deforming a source to an exemplar target shape without prespecified correspondences. It opens the opportunity for novel shape synthesis and animation with minimal human intervention, requiring only an exemplar target 3D model without any form of correspondence labels.
Such unsupervised deformation also plays an important role in improving the fitting quality of deformed CAD models to match scanned models, which is a topic of growing interest in the vision community~\cite{avetisyan2019scan2cad,avetisyan2019end,dahnert2019joint,uy2020deformation}.

\paragraph{Challenges in shape deformation} Shape deformation is a conceptually challenging problem. A good shape deformation usually requires a good alignment of the deformed shape to the target model while preserving the shape features of the source model.
The first criterion usually conflicts with the second since a perfect alignment to the target precludes the preservation of the source features.
To balance the criteria, traditional methods~\cite{sorkine2007rigid,jiang2017consistent,sorkine2009interactive} jointly optimize for the rigidity of the source mesh and the fitting error of the correspondences specified by humans.
For unsupervised shape deformation without explicit correspondences, iterative-closest-point (ICP) algorithm~\cite{besl1992method} can be used to provide supervision for the optimization process. However, it can practically yield severe artifacts given complex CAD model structures. Recent works introduce deep learning methods to directly predict point-wise offsets~\cite{wang20193dn} or to predict a control cage~\cite{yifan2019neural} for more global deformations.
However, direct predictions of vertex offsets do not offer sufficient regularization to the resulting shape, oftentimes leading to corrupt shape features such as self-intersections and non-physical distortions. On the other hand, predicting a coarse control cage adds too much regularization to the process, limiting the admissible degrees of freedom to allow a good fit between the deformed geometry and the target shape, leading to large fitting errors. Furthermore, these learning-based methods suffer from generalization issues and have limited capability for handling novel shapes.

\paragraph{Proposed formulation} We propose a novel and intuitive flow-based deformation framework that attempts to address all of the aforementioned challenges. In essence, we view the process of shape deformation as the process of convecting the geometric manifold of the source shape at time $t=0$ along with a temporally varying flow $[v_x, v_y, v_z] = f_{\theta}(x, y, z, t)$ to time $t=1$ to match the target geometry. The flow $f_{\theta}(x, y, z, t)$ can be parameterized using a neural network with trainable parameters $\theta$ that can be optimized via gradient descent. The deformation process can be integrated using a differentiable ODE solver~\cite{chen2018neural}. We use a simple geometric matching loss
(the 2-way squared Chamfer Distance computed using the ICP algorithm) between the deformed shape and the target shape together with a standard rigidity loss to optimize the neural flow model.

The flow-based formulation has various inherent advantages compared with competing approaches.
First, we show that any mesh deformation-induced via a sufficiently smooth and continuous flow is guaranteed to prevent the self-intersection of the original surfaces. This is crucial in limiting the space of deformations to a physical and visually appealing one since it is one of the main challenges facing models that directly predict vertex offsets or optimize for vertex positions. For example, we guarantee that front and back faces of table tops do not collide during deformation.
Second, we show that the flow naturally induces a bijective mapping $\mathcal{D}: \mathbb R^3 \rightarrow \mathbb R^3$ between the source and target geometries, which inherently guarantees the \emph{source to target to source} cycle consistency for deformations, which further eases the optimization when we adopt the bidirectional fitting loss.
Third, compared to a traditional parameterization of deformations using control cages or control points, the flow-based approach allows for a much higher degree of freedom for deformations due to the high representational capacity of neural networks. Last but not least, the continuous and smooth nature of the flow allows for seamless interpolation between the source and target states, allowing for smooth animations between shapes.

\paragraph{Mesh Preprocessing} To further extend our model to ``CAD meshes in the wild" (e.g. ShapeNet~\cite{chang2015shapenet}) that contain artifacts such as irregular triangles, non-uniform vertex distribution, non-watertight connectivity, topologically disconnected components, and self-intersections, we propose a robust computational pipeline for mitigating such irregularities, 
including a mesh subdivision operation, a virtual link construction, and a uniform skeleton template representation (Section~\ref{sec:represent}). These steps aim to produce uniform vertex samples with geometry-aware connections for rigidity loss modeling, and thereby guarantee a fast and valid deformation optimization and preserve global and local details (e.g. geometric connectivity and sharp features).
As a result, our ICP-based optimization does not have artifacts reported by~\cite{yifan2019neural} and significantly outperforms the existing learning approaches based on fitting errors.

We provide a benchmark consisting of randomly selected pairs of shapes from the ShapeNet repository inside the chair, table, and sofa categories, where each pair is from the same category. We exhaustively evaluate the performance of different methods. The experiment shows that we significantly outperform the state-of-the-art according to fitting errors. We demonstrate that our framework is beneficial for many downstream applications including shape interpolation and animation, scan-to-CAD fitting, and texture reconstruction.
In summary, our main contributions are:
 \begin{itemize}
 \setlength{\topsep}{0pt}
 \setlength{\parsep}{0pt}
 \setlength{\partopsep}{0pt}
 \setlength{\parskip}{0pt}
 \setlength{\itemsep}{2pt}
     \item Introduction of a novel bijective flow-based deformation function with intrinsic smoothness and powerful representation capacity.
     \item Formulating a robust mesh preprocessing pipeline for dealing with complex structures of CAD models, thus enabling a practical, robust and scalable deformation framework.
     \item Evaluating deformation performance on a benchmark dataset, in which we significantly outperform the state-of-the-art.
     \item Illustration of the applicability of our framework to benefit various downstream applications including shape animation, scan-to-CAD fitting, and texture reconstruction.
 \end{itemize}

\section{Related Works}

\paragraph{Mesh Deformation.} Detail-preserving deformation is an important direction in geometry processing and has been studied for decades~\cite{sorkine2009interactive}. In supervised deformation, users prespecify corresponding pairs of points in the source and target shape, upon which the deformation algorithm deforms the source shape to match the target on the prescribed points using some sort of interpolation~\cite{sorkine2007rigid,huang2008non,chao2010simple,levi2014smooth}. The key challenge is to ensure feature preservation by enforcing minimum distortion or rigidity. Without explicit correspondences, our deformation method optimizes the fitting and rigidity loss using correspondences from the iterative-closest-point (ICP) algorithm. Our novel neural deformation function is intrinsically smooth and does not require strong rigidity loss during optimization, which results in better fitting quality.

A different class of methods use a smooth parametric function so that the structure and the shape details are guaranteed to be preserved during the interpolation process. Such smooth functions can be implemented as the deformation of vertices on a coarse cage mesh template so that the deformation of the surface points are interpolated from the template~\cite{ju2005mean,joshi2007harmonic,lipman2008green,thiery2012cager,xian2012automatic,sacht2015nested,calderon2017bounding}. Other methods implement such functions with coarse regular or spherical grid structures to model the deformation in the image or camera space~\cite{zhou2013elastic,zhou2014simultaneous,huang20176}. However despite successes in these methods to preserve geometric details, a deformation cage has limited degrees of freedom to describe subtle deformations. Although regular grids have more degrees of freedom, they require a strong rigidity constraint to avoid severe distortion and thus also has limited capacity for deformation. Our parametric function is a neural network-based function which has powerful representation capacity for complex deformations, while its intrinsic smoothness ensures a sufficient level of rigidity. Moreover, our deformation is a bijective mapping that enables unsupervised continuous deformations between shapes from both directions.

Recent works try to incorporate learning-based methods for shape deformation. For example, \cite{groueix20183d,wang2018pixel2mesh} learns deformations of shapes based that share a common mesh template. This approach is limited by a strong assumption of common topology between the geometries that are modelled. A different set of methods model the deformation as a field stored on a 3D lattice grid~\cite{yumer2016learning,hanocka2018alignet,jack2018learning}, where the resolution limits the capacity to accurately represent the source shapes. NeuralCage~\cite{yifan2019neural} is able to preserve shape details by learning a deformation cage, but it suffers from limited representation capacity similar to the traditional cage-based deformation techniques. Finally, such deep-learning based methods are hard to generalize to shapes beyond the training set. In our proposal, the deformation function is a neural network, but we directly optimize for the parameters of the network and therefore do not suffer from generalization issues.

\paragraph{Neural Bijectors as Normalizing Flows} In the machine learning community, generative models are usually considered as a mapping from a prior distrution (e.g., a normal distribution) to a data distribution. Normalizing Flows~\cite{rezende2015variational} is proposed as a bijective map in order to effectively model a complex distribution. Bijection allows such flows to be trained by learning the mapping from the data distribution to the prior distrubition by using the maximum likelihood loss, and the inverse can be used to sample new data from the prior. Among the first proposed approaches are the planar and radial flows whose Jacobian is easy to compute. Such models lack expressivity in higher dimensions, hence autoregressive flow models are proposed to address the issues~\cite{DBLP:journals/corr/KingmaSW16,dinh2016density,papamakarios2017masked}. NeuralODE~\cite{chen2018neural,grathwohl2018ffjord} models the flow as an integration of velocity through the time, and allow continuous transformation between distributions. For example, PointFlow~\cite{yang2019pointflow} develops a continuous normalizing flow based on it for shape generation. While all existing methods train the neural network to represent map a prior distribution to a target data distribution, we consider it as a bijective map with powerful presentation capacity that directly maps between two geometries in the physical space. We formulate it as a light-weight optimization problem without the need to pretrain on a large-scale dataset, thus avioding the generalization problem.

\begin{figure*}
\includegraphics[width=\linewidth]{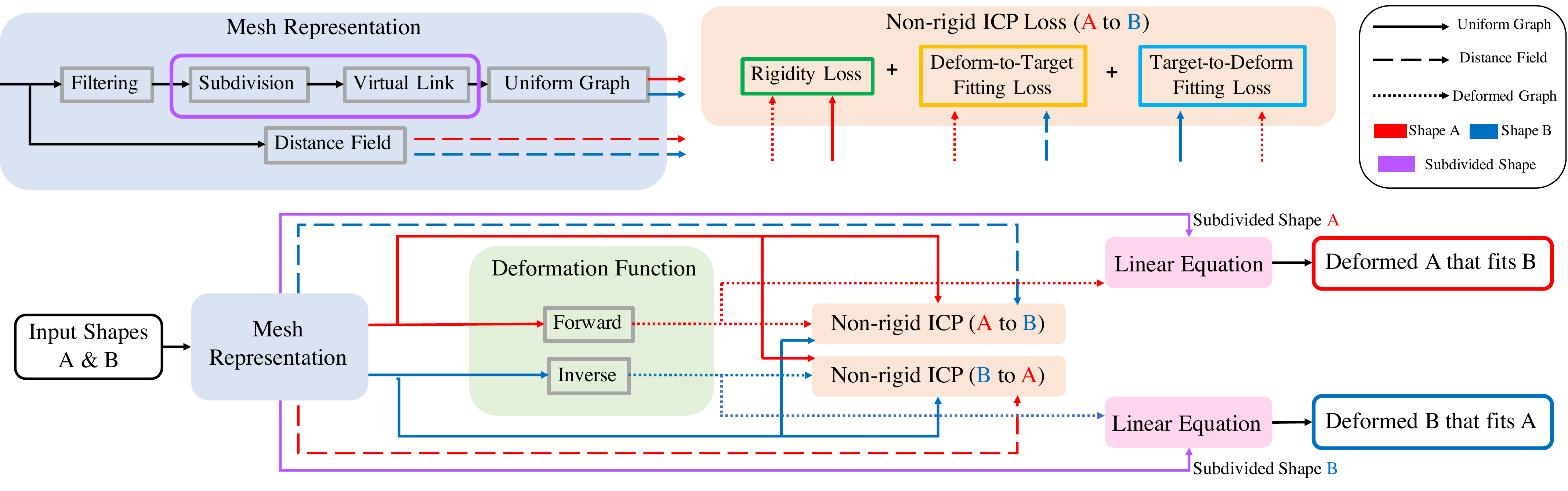}
\caption{Framework Pipeline. In the pre-processing stage, we perform a novel subdivision and build an additional uniform skeleton template. In the optimization stage, we solve for a deformation function that applies to the uniform skeleton. The optimization is guided by correspondences using iterative closest point algorithm with energy as the sum of squared distance and a rigidity loss~\cite{sorkine2007rigid}. Finally, we apply the deformation of the skeleton template to the original CAD model by solving a linear system.}
\label{fig:pipeline}
\end{figure*}
\paragraph{Shape Registration.} Our optimization process can be viewed as obtaining a deformation for minimizing the difference between two shapes. A similar topic called shape registration has been researched for decades to align point clouds from scans. Iterative closest point (ICP)~\cite{besl1992method,chen1992object} is an algorithm employed to solve this problem. An overview of the variants of this algorithm~\cite{rusinkiewicz2001efficient,diez2015qualitative} suggests that the key challenges are finding better correspondences and performing outlier-tolerant optimization. \cite{fitzgibbon2003robust} minimizes the distance between samples using the Levenberg-Marquardt solver. \cite{mitra2004registration} propose to identify the closest point efficiently using the distance field for shape optimization. Our problem is more like a non-rigid ICP problem~\cite{sumner2007embedded,li2008global,newcombe2015dynamicfusion}. The difference is that we are using neural ODE as a deep flexible template, and we deal with specific challenges of applying non-rigid ICP optimization for CAD models with complex geometry and topology.

\paragraph{Mesh Representation.} Subdivision is a common technique to upsample a coarse mesh using interpolation to obtain a more accurate geometry representation~\cite{catmull1978recursively,doo1978subdivision,dyn1990butterfly,kobbelt20003,stam1998evaluation}. In contrast to direct interpolation, we aim to robustly upsample the mesh, preserve the original geometry, and obtain right-angle triangles with edges following principal curvatures. While quadrangulation algorithms~\cite{jakob2015instant,huang2018quadriflow} produce such mesh properties, they do not guarantee the preservation of all geometry details. Instead, we simply sample grid positions for each triangle along axes of computed orientation fields and remesh each triangle with a 2D delaunay triangulation~\cite{lee1980two}.
Since CAD models have geometrically close but topologically disconnected components, several works repair models by manifold conversion~\cite{huang2018robust,chu2019repairing}. However,~\cite{huang2018robust} does not preserve sharp features and~\cite{chu2019repairing} produces unexpected shape boundaries. We do not rely on manifold topology, and simply build additional virtual links connecting close vertices to effectively model rigidity losses.
To further enhance the performance, we used the uniform skeleton template as a proxy and transfer the deformation to the source mesh using a linear solver. Similar ideas are commonly used for skinning and skeletal animation~\cite{capell2002interactive} where linear interpolation is not favored. In our scenario, the template graph is a relatively dense approximation where a linear approximation is sufficiently accurate.
\section{Overview}
\label{sec:overview}
Figure~\ref{fig:pipeline} shows a schematic of the proposed deformation framework.
In the pre-processing stage, we perform a subdivision and build a uniform skeleton template as a proxy.
In the optimization stage, we solve for a deformation function that applies to the uniform skeleton. The optimization is guided by correspondences using the ICP algorithm with the sum of squared distances together with a rigidity loss~\cite{sorkine2007rigid} as the energy to minimize.
Finally, we transfer the deformation of the skeleton template to the original CAD model by solving a linear system.
We discuss the pre-processing step in section~\ref{sec:represent} and the mathematical formulation for the deformation problem in section~\ref{sec:deform}.
We discuss the creation of the benchmark in section~\ref{sec:eval} and evaluate our methods by comparison with the state-of-the-art methods. We further perform ablation studies to study the representation capacity using different deformation functions.
Finally, we show that \ours{} benefits several downstream applications in section~\ref{sec:application}.
\section{Mesh Representation}
\label{sec:represent}
In this section, we aim at preprocessing the CAD model, so that most regions are isosceles right triangles with sufficient density to represent the deformation. Furthermore, we build virtual links and extract a uniform skeleton template as a deformation proxy for efficient optimization.

\subsection{Mesh Filtering}
Given a CAD model represented as a triangle mesh, we first normalize the mesh by applying a single translation and a uniform scale to all vertices so that the bounding box of the mesh is centered at $(0.5,0.5,0.5)$ and the maximum length along three axes is $1$.
Next, we remove degenerate elements by first looping over all triangles and removing degenerate ones containing zero area. Then, we merge duplicate vertices, such that the cleaned mesh is free from degenerate vertices, edges, and faces. 

Denote the cleaned mesh at this step as $\mathcal{\hat{M}}=\{\mathcal{\hat{V}},\mathcal{\hat{E}},\mathcal{\hat{F}}\}$ where $\mathcal{\hat{V}}=\{\mathbf{\hat{v}}_i\}$ is a set of vertices, $\mathcal{\hat{E}}=\{\mathbf{\hat{e}}_i\}$ a set of undirected edges, and $\mathcal{\hat{F}}=\{\mathbf{\hat{f}}_i\}$ a set of triangles. Specifically, $\mathbf{\hat{v}}_i$ is the 3D coordinate of the $i$-th vertex. $\mathbf{\hat{e}}_i=<\hat{e}^1_i,\hat{e}^2_i>$ is set of two vertex indices of the $i$-th undirected edge in the mesh. $\mathbf{\hat{f}}_i=<\hat{f}^1_i,\hat{f}^2_i,\hat{f}^3_i>$ is the set of three vertex indices of the $i$-th triangle in the mesh.

\subsection{Mesh Subdivision}
\begin{figure}
    \centering
    \includegraphics[width=\linewidth]{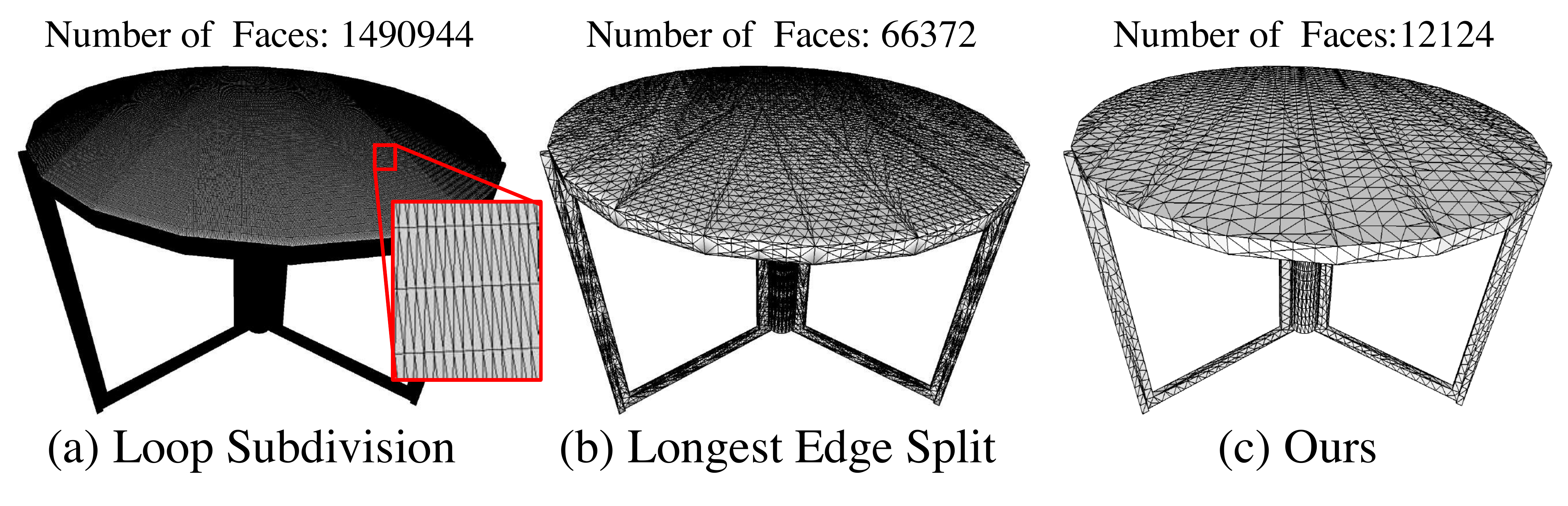}
    \caption{Subdivision results with different approaches. Classical subdivisions result in meshes with big size and contains many irregular triangles. Our subdivision leads to fewer triangles that are less skew and more regularized.}
    \label{fig:subdivision-vis}
\end{figure}
The next step involves subdividing the mesh such that the length of each edge is below a user-specified threshold: $\theta_l$.
There are multiple options for subdividing and upsampling the original mesh. For example, we can repeatedly subdivide the current longest edge and adjacent triangles until all edge lengths are below $\theta_l$. We notice that this method tends to produce very large meshes while most new triangles subdivided from an irregular triangle are still irregular, as shown in Figure~\ref{fig:subdivision-vis}(a).
Such an issue also holds with the standard loop subdivision strategy~\cite{stam1998evaluation} where at each step we subdivide each triangle into four by connecting the midpoints of the three edges (Figure~\ref{fig:subdivision-vis}(b)).
Another important thing to consider is the edge direction. Since appealing deformations are usually considered to be ones that bend along the principal curvature, novel edges created from the subdivision should ideally follow the principal curvature.
Such problems can potentially be addressed by the state-of-the-art quadrangulation methods~\cite{jakob2015instant,huang2018quadriflow}. However, such a remeshing process could change the original geometry and does not work well with disconnected components that exist in raw CAD models.

The key idea of our subdivision strategy is to upsample the mesh by sampling positions of a tangent lattice grid inside each triangle $\mathbf{\hat{f}}_i$ at the resolution of $\theta_l$, where the lattice grid is fully determined by defining its origin as $\hat{f}^1_i$ and its x-axis as the orientation field computed from~\cite{huang2018quadriflow}.
In order to connect newly sampled points across different triangles sharing edges, we additionally subdivide each edge $\mathbf{\hat{e}}_i$ uniformly into segments such that the length of each segment is below $\theta_l$. The number of segments can be determined as $N(\mathbf{\hat{e}}_i) = \ceil{|\mathbf{v}_{\hat{e}^1_i} - \mathbf{v}_{\hat{e}^2_i}|/\theta_l}$.
Finally, each triangle can be remeshed by running a 2D Delaunay triangulation~\cite{lee1980two} algorithm to connect interior sampled grid positions and endpoints of the segments at the triangle boundary.
As shown in Figure~\ref{fig:subdivision-vis}(c), our subdivision algorithm leads to fewer and more regular triangles. Additionally, internal edges follow principal curvatures which enable appealing bending deformations.
In the following sections, we denote our subdivided mesh as $\mathcal{M}=\{\mathcal{V},\mathcal{E},\mathcal{F}\}$.

\subsection{Virtual Links}
\begin{figure}
    \centering
    \includegraphics[width=\linewidth]{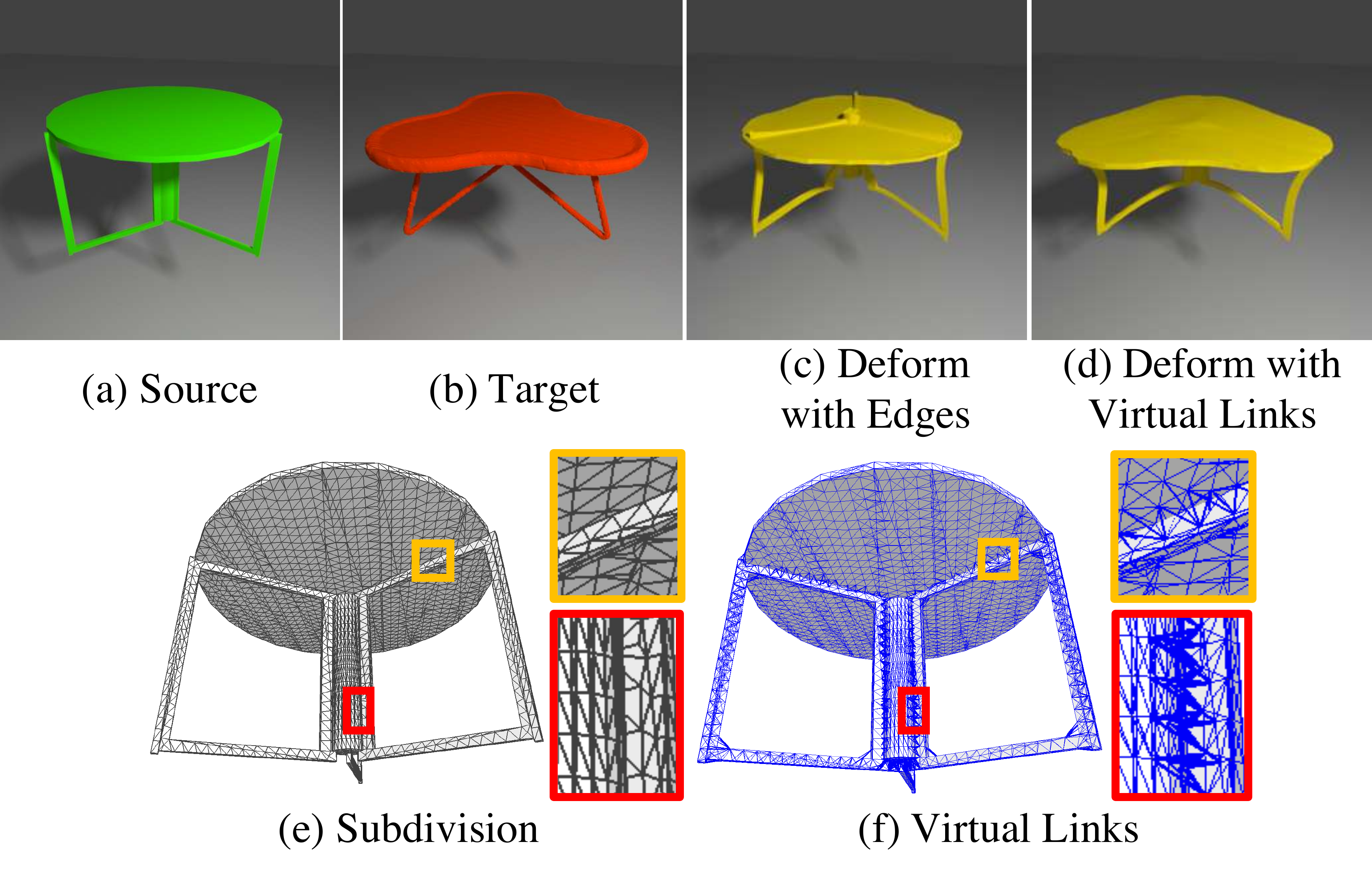}
    \caption{Deformation directly using mesh edges breaks the geometry in (c), since CAD models do not guarantee watertight connectivities. In (d), deformation with virtual links addresses this issue. Virtual links additionally connect different table parts in (f) compared to mesh edges in (e).}
    \label{fig:virtuallink-vis}
\end{figure}
Since CAD models created by artists do not guarantee watertight connectivity, it is common that different object parts are topologically disconnected while geometrically self-intersecting (e.g. table tops and legs). As a result, deformation based merely on topological connectivities from $\mathcal{E}$ breaks the geometry.
Therefore, we need to additionally consider geometric relationships between vertices.
We adopt a simple criterion that two vertices in $\mathcal{V}$ are geometrically related if their Euclidean distance is below a threshold $\theta_d$.%
However, explicitly storing such geometrical relationships for all pairs of close vertices is quite inefficient. Specifically, the number of geometrical relationships is quadratic to the number of vertices in the local region, while some of them can have a large number of vertices in the CAD models.

In our implementation, we create a sparse 3D uniform voxel grid with the resolution as $\theta_d$ in a unit cube, where each non-empty voxel contains a list of vertices inside the voxel.
We loop over each non-empty voxel, collect the vertices from this voxel and its subsequent three voxels along the three axes.
We run a 3D Delaunay triangulation to connect the vertices and keep the edges whose lengths are smaller than $\theta_d$. Such a formulation ensures the number of constraints is with $O(N)$ complexity in the local region.
We collect all these edges together with $\mathcal{E}$ as a set of virtual links $\mathcal{L}$. We use the virtual links for enforcing the rigidity constraints, such that the deformation guarantees local geometrical feature preservation without a requirement for watertight manifold topology.This is illustrated in the comparison between Figure~\ref{fig:virtuallink-vis}(c) and (d). Virtual links visualized in Figure~\ref{fig:virtuallink-vis}(f) additionally connect different table parts in addition to mesh edges, as can be seen in Figure~\ref{fig:virtuallink-vis}(e).

\subsection{Uniform Skeleton Template}
It is sufficient to solve the deformation for $\mathcal{V}$ with given constraints $\mathcal{L}$ such that the deformed mesh can be represented as the updated vertex positions connected by the triangles $\mathcal{F}$. However,though the upper bound of edge lengths through the previous steps can be guaranteed, the distribution of vertices is dense and non-uniform, rendering the optimization process inefficient. Short edges can also cause numerical issues and thereby require small time steps during optimization.

Therefore, we additionally build a uniform skeleton template as a graph to approximate the original mesh graph. Again, we create a sparse 3D uniform voxel grid at the resolution of $\theta_g$ and for each voxel we collect the vertices from $\mathcal{V}$ that belong to it. For each non-empty voxel, we create a node as the average of the vertices within. We define the set of created nodes as $\mathcal{V}_G$. For each virtual link in $\mathcal{L}$, we create a graph edge connecting to the corresponding node that vertices of the link belong to. Note that several virtual links correspond to the same edge in the graph template, and a single edge is kept for these links. We define the set of graph edges as $\mathcal{E}_G$. Finally, for the $i$-th graph node $\mathbf{v}_i\in \mathcal{V}_G$, we store the set of vertex indices in the original mesh $\mathcal{V}$ as $s_i$, and define the collection of $s_i$ as $\mathcal{S}$. We denote such a uniform skeleton template as $\mathcal{G} = \{\mathcal{\tilde{V}}, \mathcal{\tilde{E}}, \mathcal{S}\}$.
\section{Mesh Deformation}
\label{sec:deform}
We aim at solving for the deformation map between a pair of two meshes A and B, preprocessed via the proposed approach in Sec.~\ref{sec:represent}, that includes the subdivided vertices, triangles, virtual links and skeleton template. Specifically, we denote the two meshes as $\mathcal{M}_A=\{\mathcal{V}_A,\mathcal{F}_A,\mathcal{L}_A,\mathcal{G}_A\}$ and $\mathcal{M}_B=\{\mathcal{V}_B,\mathcal{F}_B,\mathcal{L}_B,\mathcal{G}_B\}$, where the skeleton templates are $\mathcal{G}_A=\{\mathcal{\tilde{V}}_A,\mathcal{\tilde{E}}_A,\mathcal{S}_A\}$ and $\mathcal{G}_B=\{\mathcal{\tilde{V}}_B,\mathcal{\tilde{E}}_B,\mathcal{S}_B\}$.

\subsection{Energy Minimization}
We solve for the deformation in two steps. First, we deform the source skeleton template to the target shape by minimizing fitting and rigidity energies. Second, we solve a linear system to transfer the deformation of the skeleton template to the source mesh.

Suppose shape A is the source and B is the target, the energy is as follows:
\begin{equation}
    E(\mathcal{M}_A,\mathcal{M}_B,\mathcal{D}) = E_D(\mathcal{D}(\mathcal{G}_A),\mathcal{M}_B) + \lambda \cdot E_R(\mathcal{D}(\mathcal{\tilde{V}}_A),\mathcal{\tilde{V}}_A,\mathcal{\tilde{E}}_A).
    \label{eq:energy}
\end{equation}
The energy contains the fitting loss $E_D$ (Section~\ref{sec:fitting}) between the deformed skeleton template $\mathcal{D}(\mathcal{\tilde{G}}_A)$ and the target mesh $\mathcal{M}_B$, and the rigidity loss $E_R$ (Section~\ref{sec:rigidity}) of the deformed skeleton template $\mathcal{D}(\mathcal{\tilde{G}}_A)$, given the source template $\mathcal{\tilde{G}}_A$.

Furthermore, in the case of solving for the bijective deformation mapping between A and B such that either of them can be the source or target, we require $\mathcal{D}$ to be a bijective mapping and propose the the two-way deformation energy to be:
\begin{equation}
    E_2(\mathcal{M}_A,\mathcal{M}_B,\mathcal{D}) = E(\mathcal{M}_A,\mathcal{M}_B,\mathcal{D}) + E(\mathcal{M}_B,\mathcal{M}_A,\mathcal{D}^{-1}).
    \label{eq:energy2}
\end{equation}
In either scenario, we optimize $\mathcal{D}$ to minimize the energy, and obtain the deformed skeleton $\mathcal{D}(\mathcal{G}_A)$ and optionally $\mathcal{D}^{-1}(\mathcal{G}_B)$. In the post processing step, we solve a linear system to apply the deformation to the original mesh (Section~\ref{sec:linear}).

\subsection{Deformation Function}
\label{sec:deform-func}
The deformation function $\mathcal{D}: \mathbb R^3 \rightarrow \mathbb R^3$ maps every point in the source shape to the deformed shape. 

First, our framework can incorporate traditional methods~\cite{sorkine2007rigid} and explicitly represent $\mathcal{D}$ as the set of sampled vertex coordinates in the deformed mesh, which we denote as $\mathcal{D}_e$, in which case the variables we are optimizing are the  corresponding locations of these vertices in the deformed source skeleton, initialized as $\mathcal{\tilde{V}}_A)$.
Using the mesh preprocessing pipeline (detailed in Section~\ref{sec:represent}), such an optimization process effectively preserves local geometry details for complex CAD models. However, we can only use such a functions for the A-to-B deformation problem by minimizing Equation~\ref{eq:energy}, since there is no guarantee that $\mathcal{D}_e$ is invertible.

\begin{figure}
    \centering
    \includegraphics[width=\linewidth]{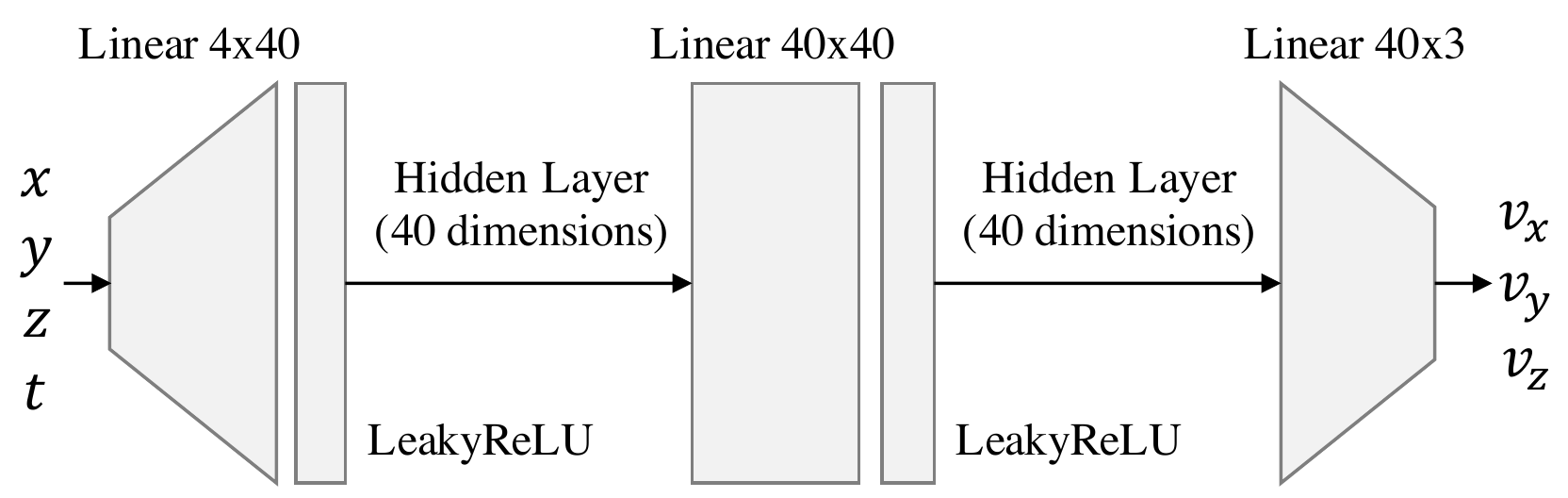}
    \caption{Neural network architecture for velocity field prediction.}
    \label{fig:network}
\end{figure}

To find an invertible deformation function by minimizing the energy in Equation~\ref{eq:energy2}, the key is to find a parameterized function $\mathcal{D}$ which is flexible to handle complex deformation while ensuring invertibility. Neural ODE~\cite{chen2018neural} satisfies both of these requirements. We first define the velocity field $\mathbf{u}(\mathbf{p},t)$ given a 3D location $\mathbf{p}$ and time $t$. The field is parameterized by a neural network architecture shown in Figure~\ref{fig:network}. Then, we create the deformation path $\mathbf{p}(\mathbf{x},t)$ for each source point $\mathbf{x}$ where instant velocity at time $t$ is $\mathbf{u}(\mathbf{p},t)$.
$\mathcal{D}_{\mathrm{ode}}$ maps all source points at time $0$ to deformed points at time $1$.
The mathematical definition of $\mathcal{D}_{\mathrm{ode}}$ is shown in Equation~\ref{eq:neural-ode}, where $\mathbf{u}$ is considered as the parameters of $\mathcal{D}_{\mathrm{ode}}$.
\begin{align}
    \mathbf{p}(\mathbf{x},0) &= \mathbf{x} \nonumber\\
    \frac{d \mathbf{p}}{d t} &= \mathbf{u}(\mathbf{p},t) \nonumber\\
    \mathcal{D}_{\mathrm{ode}}(\mathbf{x};\mathbf{u}) &= \mathbf{p}(\mathbf{x},1).
    \label{eq:neural-ode}
\end{align}

The inverse function $\mathcal{D}^{-1}_{\mathrm{ode}}$ can be inferred as
\begin{align}
    \mathcal{D}^{-1}_{\mathrm{ode}}(\mathbf{x};\mathbf{u}) &= \mathcal{D}_{\mathrm{ode}}(\mathbf{x};\mathbf{w}) \nonumber\\
    \mathbf{w}(*,t) &= -\mathbf{u}(*,1-t)
\end{align}
We use the implementation provided by Neural ODE~\cite{chen2018neural} to integrate $\mathbf{u}$ and obtain $\mathcal{D}_{\mathrm{ode}}$ and $\mathcal{D}^{-1}_{\mathrm{ode}}$ in a differentiable way. This allows us to optimize the parameters in $\mathbf{u}$ to achieve optimal deformation. In Section~\ref{sec:animation}, we show smooth intermediate deformation results as an animation by computing $\mathbf{p}(\mathcal{V}_A,t)$ at different time $t\in[0,1]$.

\subsection{Fitting Loss}
\label{sec:fitting}
We consider the fitting loss between the deformed source shape and the target shape as the sum of the squared distances between vertices in the deformed source and the respective nearest vertex in target, which is similar to the ICP algorithm~\cite{besl1992method}. Specifically, the fitting loss between a deformed skeleton $\mathcal{G}^*_A=\{\mathcal{\tilde{V}}^*_A,\mathcal{\tilde{E}}_A,\mathcal{S}_A\}$ and the target mesh $\mathcal{M}_B$ is defined as:
\begin{equation}
    E_D^a(\mathcal{G}^*_A,\mathcal{M}_B) = \sum_{\mathbf{v}^*_i\in \mathcal{\tilde{V}}^*_A} \min_{\mathbf{p}_i\in \mathcal{M}_B} ||\mathbf{v}^*_i - \mathbf{p}_i||_2^2.
    \label{eq:dis-forward}
\end{equation}
Since the target mesh $\mathcal{M}$ is a fixed shape, we precompute a distance field as a $64^3$ uniform grid in a unit cube for fast retrieval of the nearest distance. By optimizing Equation~\ref{eq:dis-forward}, we are able to obtain the deformation such that vertices in the deformed source skeleton are well-covered by the target mesh.
Since we additionally require the target mesh to be covered by the source skeleton, we define a fitting loss between the target skeleton vertices $\mathcal{\tilde{V}}_B$ of the mesh $\mathcal{M}_B$ and the deformed skeleton $\mathcal{G}^*_A$ as:
\begin{equation}
    E_D^b(\mathcal{G}^*_A,\mathcal{M}_B) = \sum_{\mathbf{v}_i\in \mathcal{\tilde{V}}_B} \min_{\mathbf{p}^*_i\in \mathcal{G}_A} ||\mathbf{v}_i - \mathbf{p}^*_i||_2^2.
    \label{eq:dis-backward}
\end{equation}
Since the source skeleton vertices are changed during optimization, we rebuild a KDtree for the deformed skeleton at every optimization step for efficiently querying the nearest distance between the target and the deformed source skeleton.

We define $E_D=E_D^a$ as a baseline (``ours-1way'') in Section~\ref{sec:eval-compare}, and define $E_D=E_D^b$ for scan-to-cad fitting in Section~\ref{sec:scantocad} since it is unreasonable to cover all regions of the CAD by a partial scan. For all other experiments, we optimize the two-way fitting loss as $E_D=E_D^a+E_D^b$.

\subsection{Rigidity Loss}
\label{sec:rigidity}
Our rigidity loss $E_R$ follows the definition in~\cite{sorkine2007rigid}. Specifically $E_R$ measures the local distortions of the deformed point set $\mathcal{V}^*$ from the original point set $\mathcal{V}$ by aggregating the neighborhood information from an edge set $\mathcal{E}$. Since the edges are undirected, $<j,i>$ is equivalent to $<i,j>$. The rigidity energy $E_R$ is defined in Equation~\ref{eq:rigidity}.
\begin{equation}
    E_R(\mathcal{V}^*,\mathcal{V},\mathcal{E}) = \min_{\{\mathbf{R}_i\}} \sum_{<i,j>\in \mathcal{E}} ||(\mathbf{v}^*_i-\mathbf{v}^*_j) - \mathbf{R}_i(\mathbf{v}_i-\mathbf{v}_j)||_2^2
    \label{eq:rigidity}
\end{equation}
In order to optimize the rigidity loss, we use a three-channel Euler angle for each vertex $i$ to represent its local rotation $\mathbf{R}_i$, and initialize it to be zero. These Euler angles are jointly optimized with the deformation function parameters.

\subsection{Post processing}
\label{sec:linear}
Once we obtain the deformation of the skeleton template, we can apply the deformation to the original mesh. First, we estimate the local rotation of the local region for each vertex as the optimized rotation of the nearest skeleton node. For $\mathcal{M}=\{\mathcal{V},\mathcal{F},\mathcal{L},\mathcal{G}\}$, we deform vertices $\mathcal{V}$ to $\mathcal{V}^*$ to minimize the following energy:
\begin{align}
    E_L(\mathbf{V}^*;\mathcal{M},\mathcal{D}) &= \sum_{i} ||\mathcal{D}(\mathbf{\tilde{v}}_i) - \frac{1}{|s_i|}(\sum_{j\in s_i} \mathbf{v}^*_j)||_2^2 \nonumber\\
    &+ \lambda \sum_{<i,j>\in \mathcal{L}} ||(\mathbf{v}^*_i - \mathbf{v}^*_j) - \mathbf{R}_i (\mathbf{v}_i - \mathbf{v}_j)||_2^2
    \label{eq:linear}
\end{align}
The first term requires that the average of vertices in $s_i$ that are close to the $i$-th skeleton node is close to the deformed node after the deformation. The second term is the rigidity loss that the deformed edge offset should be consistent with the estimated local rotation of the vertices (similar to Equation~\ref{eq:rigidity}). Since $E_L$ is quadratic to $\mathcal{V}^*$, we can solve the massive number of vertices in the original CAD models efficiently with a linear solver.

\subsection{Implementation details}
\label{sec:setup}
We set $\theta_l=0.02$ for subdivision, $\theta_d=0.015$ for building virtual links, and $\theta_g=0.01$ as the resolution of the skeleton node. We set $\lambda=1$ for Equation~\ref{eq:rigidity} when optimizing $\mathcal{D}_E$ and $\lambda=0.1$ when optimizing $\mathcal{D}_{\mathrm{ODE}}$. We adopt weaker rigidity loss since $\mathcal{D}_{\mathrm{ODE}}$ is intrinsically smooth and easier to preserve shape features.

We solve $\mathcal{D}_E$ with the efficient Levenberg-Marquardt algorithm~\cite{more1978levenberg} implemented in the Ceres Solver\footnote{\url{http://ceres-solver.org}}. We solve parameters in $\mathcal{D}_{\mathrm{ODE}}$ with an Adam solver~\cite{kingma2014adam} implemented in PyTorch\footnote{\url{https://pytorch.org}}. In the next section, we evaluate the difference between all three versions and compare them with the state-of-the-art methods in a controlled experiment.

\section{Evaluation}
\label{sec:eval}
We create our benchmark dataset by randomly selecting 3625 pairs of shapes as sources and targets from ShapeNet. Our pairs are selected from the chair, the table, and the sofa categories and we guarantee that the two shapes from each pair belong to the same category.
In Section~\ref{sec:eval-compare}, we evaluate our deformation framework and compare them with the state-of-the-art methods. We additionally do ablation studies for different mesh representations in Section~\ref{sec:eval-represent} and different deformation functions in Section~\ref{sec:eval-deformfunc}.

\subsection{Comparison of Deformation Quality}
\label{sec:eval-compare}
\begin{figure*}
\includegraphics[width=\linewidth]{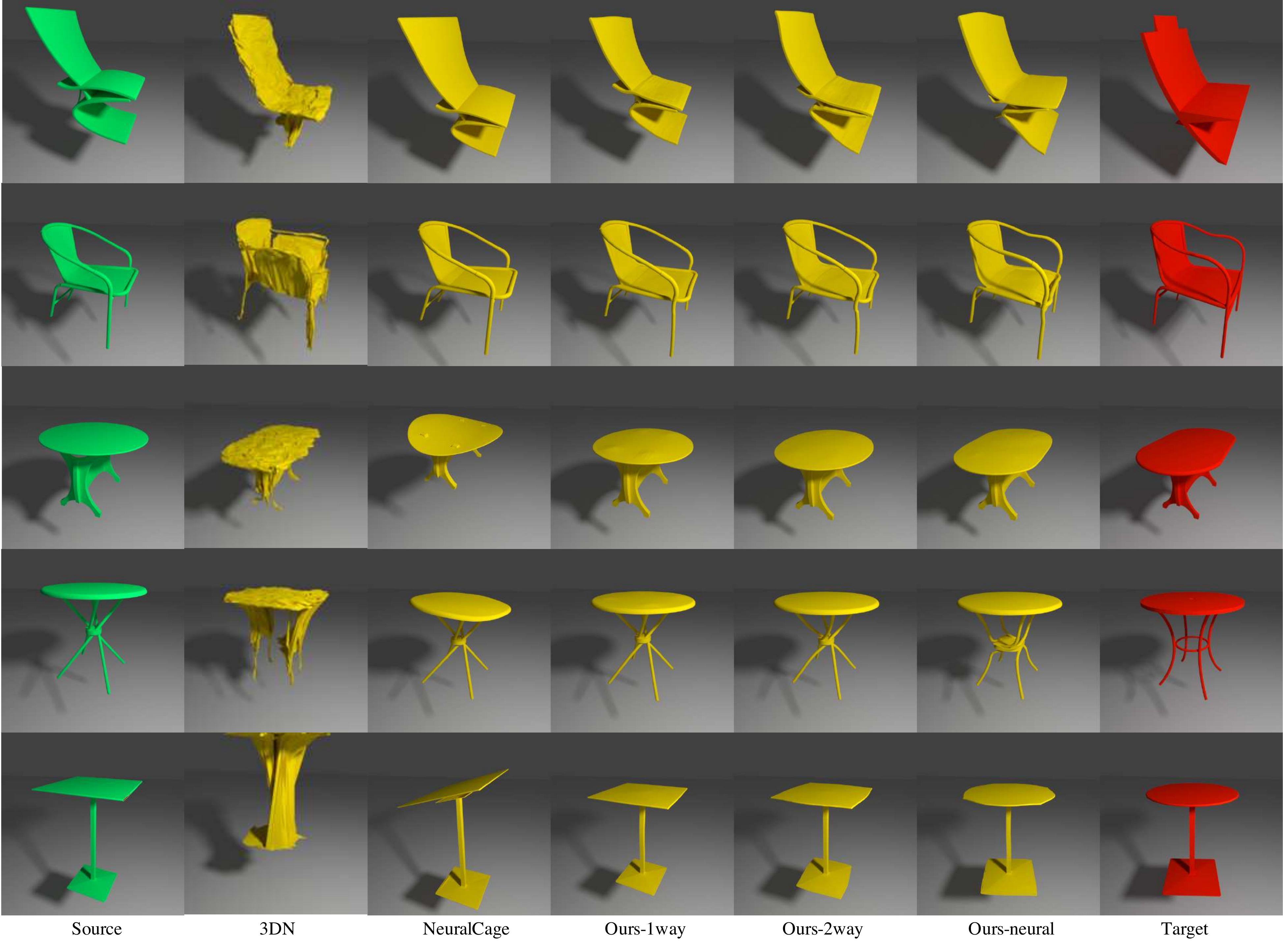}
\caption{Visual comparison on shape deformation between different approaches. 3DN~\cite{wang20193dn} and NeuralCage~\cite{yifan2019neural} do not fit the target as good as our framework do. Ours-1way is not optimized for the coverage of the target, while ours-2way ensures coverage of both the deformed and the target shape. Ours-neural also ensures the coverage given by the bijective mapping.}
\label{fig:visual-shapenet}
\end{figure*}
\begin{table}
    \centering
    \tabcolsep=0.07cm
    \begin{tabular}{|c|c|c|c|c|}
        \hline
         & Chair & Table & Sofa & All \\
         \hline
         Source & 6.232/6.232 & 6.580/6.580 & 4.805/4.805 & 6.155/6.155 \\
         \hline
         NeuralCage & 5.781/7.396 & 5.760/21.36 & 3.542/29.60 & 5.381/17.64\\
         \hline
         3DN & 3.000/7.523 & 3.325/10.843 & 2.952/9.482 & 3.489/10.44\\
         \hline
         Ours-1way & 2.914/2.914 & 2.990/2.990 & 2.622/2.622 & 2.901/2.901 \\
         \hline
         Ours-2way & 2.053/2.053 & 2.073/2.073 & 1.990/1.990 & 2.052/2.052\\
         \hline
         Ours-neural & \textbf{1.136/1.136} & \textbf{1.297/1.297} & \textbf{1.092/1.092} & \textbf{1.203/1.203}\\
         \hline
    \end{tabular}
    \caption{Evaluation using the two-way Chamfer distance ($\times 10^{-2}$). For each entry, we report two numbers, one where the shapes are canonically aligned with the training data and the other with a 90-degree rotation for both the source and the target around the the z-axis.
    }
    \label{tab:chamfer}
\end{table}
We compare our framework with the state-of-the-art methods based on the two-way Chamfer distance between the deformed source model and the target model
The state-of-the-art methods include 3DN~\cite{wang20193dn} and NeuralCage~\cite{yifan2019neural}. Our framework enables valid CAD deformation by optimizing Equation~\ref{eq:energy} using $\mathcal{D}_E$ with traditional one-way ($E_D=E_D^a$) or two-way ($E_D=E_D^a+E_D^b$) squared Chamfer loss, and our newly proposed neural deformation $\mathcal{D}_{\mathrm{ODE}}$ with two-way squared Chamfer loss. They are shortened as ``ours-1way'', ``ours-2way'' and ``ours-neural'' respectively. Note that all these experiments are performed under a single-direction scenario (Equation~\ref{eq:energy}), while we additionally discuss bidirectional optimization (Equation~\ref{eq:energy2}) in Section~\ref{sec:eval-property} and utilize use it in Section~\ref{sec:animation}.

In Table~\ref{tab:chamfer}, we report two numbers for each entry, one with the shape canonically aligned with the training data and the other with a 90-degree rotation for both the source and the target around the z-axis. All three versions of ours significantly outperform the existing state-of-the-art methods according to the fitting error. Further, the drop in performance for the deep-learning-based baselines show that they do not generalize to novel shapes or poses outside the training regime with a simple rotation variation.
Visual comparisons are shown in Figure~\ref{fig:visual-shapenet}:
Our method achieves better fitting quality and visual appeal compared to 3DN~\cite{wang20193dn} and NeuralCage~\cite{yifan2019neural}. Ours-1way is not optimized for the coverage of the target, while ours-2way ensures coverage of both the deformed and the target shape. Ours-neural additionally provides more flexible deformation with better fitting quality.
Based on the experiments, our optimization-based methods significantly outperform the existing learning-based methods with respect to the fitting errors, and our novel neural function achieves the best performance.

\begin{table}
    \centering
    \tabcolsep=0.08cm
    \begin{tabular}{|c|c|c|c|c|}
        \hline
         NeuralCage & 3DN & Ours-1way & Ours-2way & Ours-neural\\
         \hline
          1.17/4.81 & 0.58/2.12 & -/1.24 & -/3.79 & 3.73/3.96\\
         \hline
    \end{tabular}
    \caption{Average time (seconds) used to process each pair of shapes on a server with 8 Titan X GPUs or 32 2.20GHz CPUs. For NeuralCage, 3DN and ours-neural, we report two numbers, running it with or without GPUs.}
    \label{tab:methods-perf}
\end{table}
Efficiency is another important factor to consider. We process all models on a server with 8 Titan X GPUs and 32 2.20GHz CPUs. The average processing time for each pair of shapes for different methods is shown in Table~\ref{tab:methods-perf}. Regarding efficiency, we find that our optimization based method: $\mathcal{D}_{E}$ is on par with the forward pass of a trained network (NeuralCage) which relies on GPUs to do efficient computation with a big network. Our NeuralODE-based optimization uses a significantly more lightweight neural network which is easy to optimize. The gap between GPU and CPU of our NeuralODE is small, due to our reliance on a CPU implementation of Equation~\ref{eq:energy}. This suggests that our GPU version has the potential to be further accelerated. Finally, we directly obtain a bijective and smooth intermediate deformation animation (Section~\ref{sec:animation}) with infinite time resolution once our single-pass optimization is done.

\subsection{Choice of Mesh Representation}
\label{sec:eval-represent}
\begin{table}
    \centering
    \tabcolsep=0.12cm
    \begin{tabular}{|c|c|c|c||c|c|}
        \hline
         & Loop & Longest & Our-CAD & Watertight & Skeleton\\
         \hline
         Error ($\times 10^{-2}$) & 2.043 & 2.286 & 1.925 & 1.823 & 2.052\\
         \hline
         Time (s) & 141.7 & 51.6 & 7.14 & 5.73 & 3.79\\
         \hline
    \end{tabular}
    \caption{Performance comparison for different mesh representations. ``Loop" stands for loop subdivision~\cite{stam1998evaluation}. ``Longest" stands for longest-edge subdivision. ``Our-CAD" uses our subdivision method. ``Watertight" uses ~\cite{huang2018robust} to convert the CAD model into a watertight shape. ``Skeleton" uses our uniform skeleton approximation based on our subdivision strategy.}
    \label{tab:template-perf}
\end{table}
\begin{figure}
    \centering
    \includegraphics[width=\linewidth]{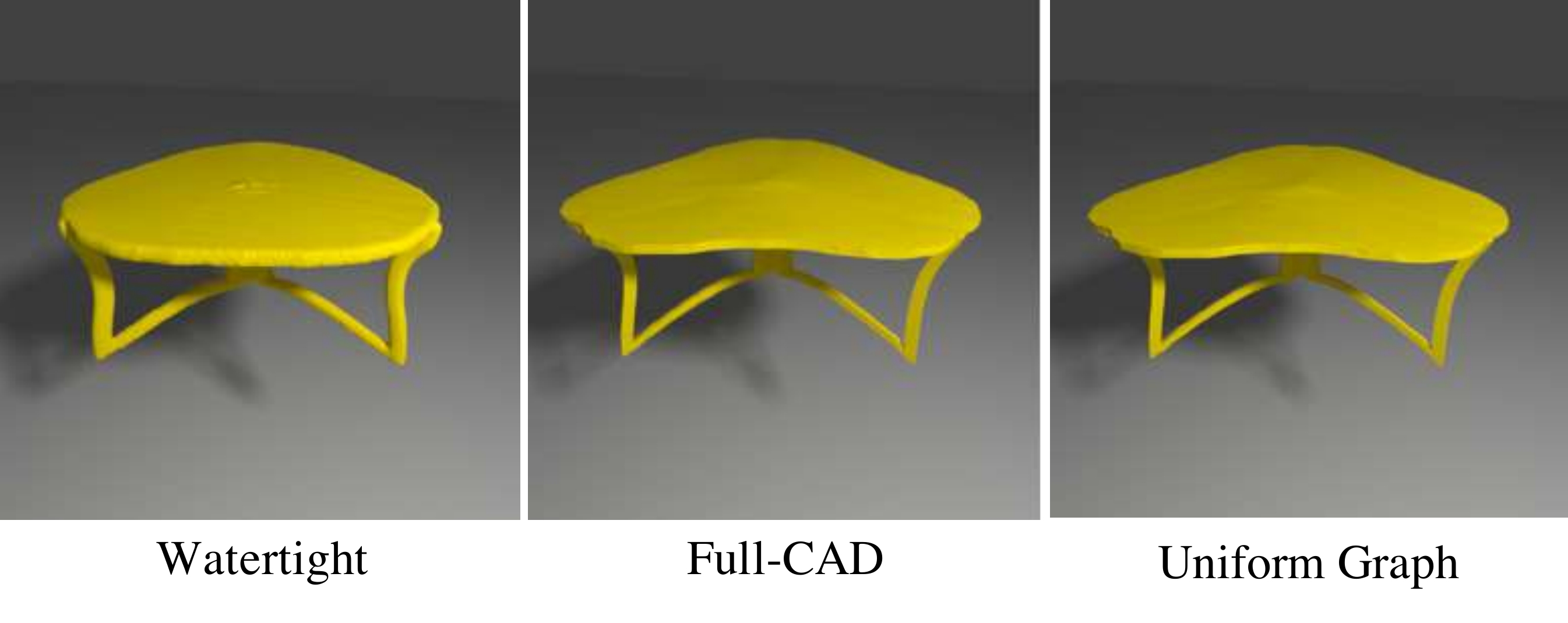}
    \caption{Deformation results with different mesh representations. Although watertight approximation~\cite{huang2018robust} can achieve good fitting to the target, the geometry is different and sharp features can not be well-preserved. Our skeleton template representation is a good approximation which generates quite similar results comparing with direct optimization of the original CAD model.}
    \label{fig:template-vis}
\end{figure}
 In this ablation study, we investigate the deformation performance based on different mesh representations  obtained through different mesh preprocessing pipelines. We optimize Equation~\ref{eq:energy} for all variants using our proposed method and report the fitting error and timing in Table~\ref{tab:template-perf}.
Direct optimization of the original CAD model pre-processed using our subdivision method achieves the minimum fitting error.  Other subdivision methods achieve slightly worse fitting quality because of the irregulararity of the triangles. Although the watertight approximation~\cite{huang2018robust} achieves good fitting to the target, the geometry is different and the sharp features can not be well-preserved, as shown in figure~\ref{fig:template-vis}. The skeleton template representation achieves similar fitting error comparing 
to other subdivision methods.
Efficiency-wise, deformation based on our subdivision is much faster than existing ones. Our skeleton template further improves efficiency.
To balance quality and efficiency, we combine our subdivision method and the skeleton template for mesh representation.

\subsection{Choice of Deformation Function}
\label{sec:eval-deformfunc}
\begin{figure}
    \centering
    \includegraphics[width=\linewidth]{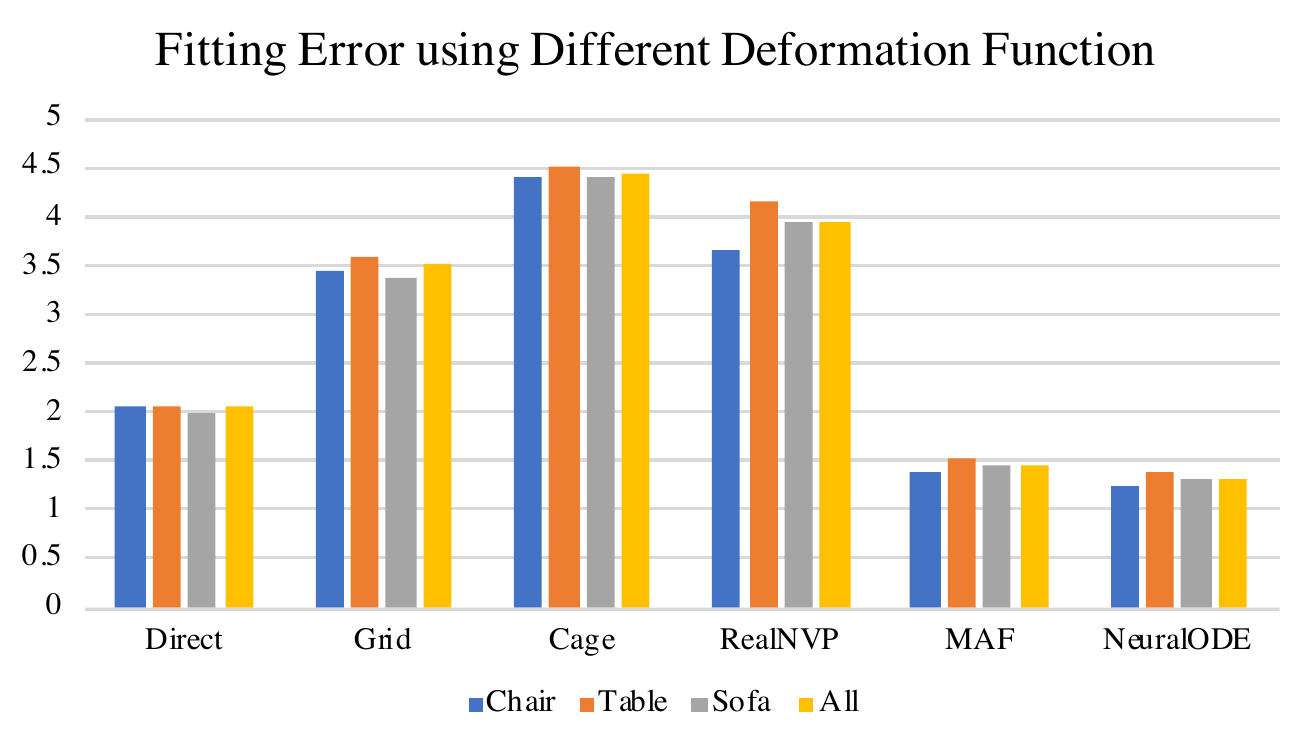}
    \caption{Representation capacity using different deformation functions. We optimize Equation~\ref{eq:energy} for all deformation functions. We set rigidity loss as $1$ for $\mathcal{D}_e$ and $\mathcal{D}_{\mathrm{grid}}$ and $0.1$ for other methods. $\mathcal{D}_{\mathrm{ode}}$ has the closest representation capacity to $\mathcal{D}_e$ and require small rigidity and thus achieves the best results comparing to other neural-network based methods. $\mathcal{D}_e$ and regular grid requires a big rigidity loss to avoid unpleasant distortion, which also limits their representation capacities. Cage-based deformation is more global and is not suitable for aligning subtle geometry structures.}
    \label{fig:deform-func}
\end{figure}
 In this ablation study, we study the impacts of different choices of deformation functions on fitting errors. For traditional deformation functions, we test $\mathcal{D}_e$ along with two common choices for parameterizing the deformation space, including uniform grids or control cages. We use the $10^3$ grids with 0.1 resolution and the convex hull of the source shape as the control cage.  For learned bijective mapping, we explore the usage of several flow-based models including RealNVP~\cite{dinh2016density}, MAF~\cite{papamakarios2017masked}, and our $\mathcal{D}_{\mathrm{ode}}$.

To test the representation capacity, we optimize Equation~\ref{eq:energy} for all deformation functions. Based on Figure~\ref{fig:deform-func}, deformation grids and cages have limited representation capacity compared to $\mathcal{D}_e$. $\mathcal{D}_{\mathrm{ode}}$ is the best choice among flow-based models and achieves better fitting quality by requiring less explicit rigidity constraints.

\subsection{Part-aware Deformation}
\begin{figure}
    \centering
    \includegraphics[width=\linewidth]{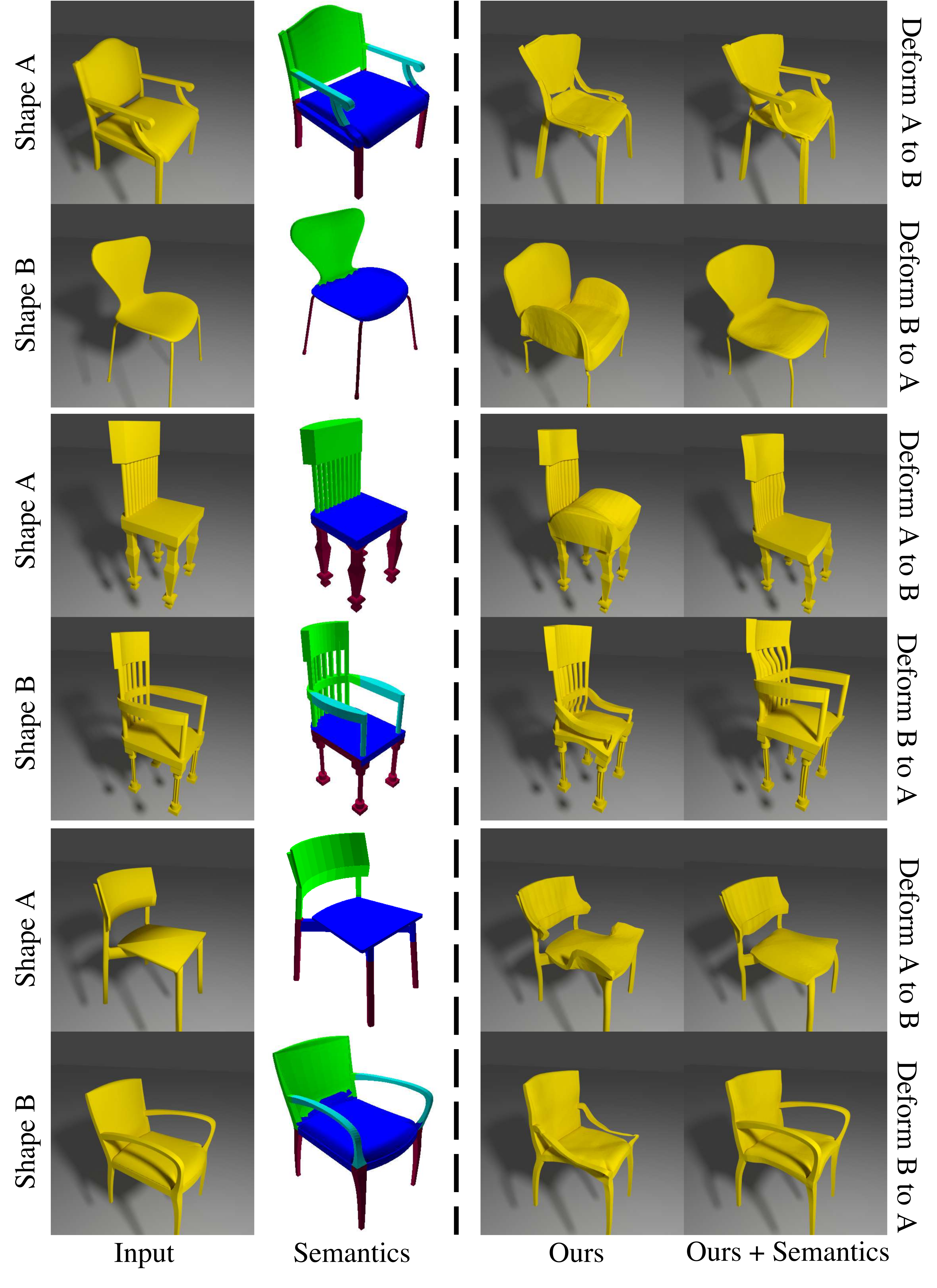}
    \caption{Part-aware deformation. Each example contains two rows as a pair of deformations. Our input are meshes (first column) with part semantics predicted by PointNet~\cite{qi2017pointnet} (second column). The third column is the pairwise deformation with our NeuralODE by optimizing the pure geometry fitting loss using Equation~\ref{eq:energy2}. We can obtain semantically meaningful deformation (fourth column) by measuring fitting loss separately for different semantic parts.}
    \label{fig:semantic}
\end{figure}
Furthermore, we explore the use of additional semantic information in our framework during pairwise shape deformations. Our inputs are pairs of chairs with semantic segmentation labels produced using PointNet~\cite{qi2017pointnet}. We separately measure the fitting loss (Section~\ref{sec:fitting}) for regions with different semantic labels and compute the sum of them as the total fitting loss. For certain semantic parts of one object that does not exist in another, we set the fitting loss as zero. In Figure~\ref{fig:semantic}, we show that such a modification can apply semantically meaningful deformations (e.g., chair handles are not deformed to seats).

\subsection{Properties of \ours{}}
\label{sec:eval-property}
We derive and evaluate several properties of \ours{}.

\paragraph*{Bijection} If $\mathcal{D}$ is a bijective mapping, we expect it deforms any point $\mathbf{p}$ to a new location, while its inverse maps it back to the same point. We measure quality of bijection for a point $\mathbf{p}$ as
\begin{equation}
\epsilon(\mathbf{p}) = ||\mathbf{p}-\mathcal{D}^{-1}(\mathcal{D}(\mathbf{p}))||
\end{equation}
From our numerical evaluation, the maximum and the average of $\epsilon(\mathbf{p})$ for all mesh vertices in the dataset are $3\times 10^{-5}$ and $6\times 10^{-7}$ respectively, compared to the mesh vertex coordinate range from 0 to 1. Therefore, the numerical errors do not affect the bijectivity of our model.

\begin{figure}
\includegraphics[width=\linewidth]{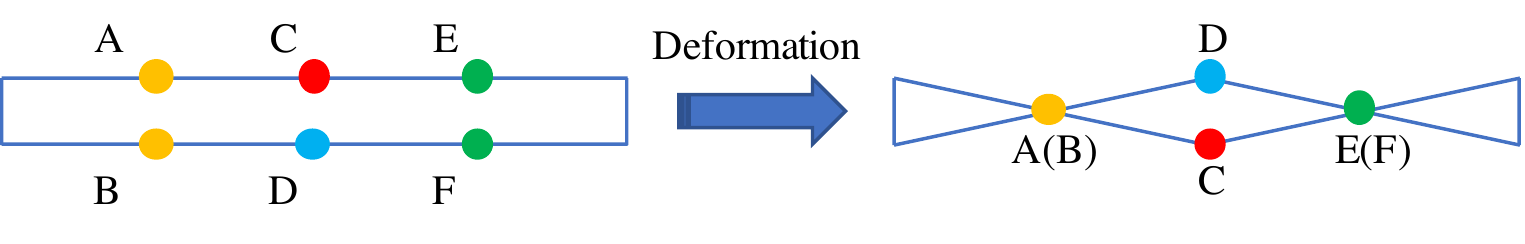}
\caption{Schematics for the deformation of points on a tabletop. If the deformation maps two different points $A$ and $B$ to the same location and causes self-intersections, such deformation would violate the property of bijection, therefore bijective mappings are free of self-intersections.}
\label{fig:bijection}
\end{figure}
\paragraph*{Free of Self-Intersection} An intuition based illustration of proof by contradiction is given in Figure~\ref{fig:bijection}. Any deformation achieved through a bijective map does not induce self-interesction of any two points in the original space. This is important for dealing with thin structures (e.g. tabletop) without enforcing strong rigidity constraints. 

\begin{figure}
\includegraphics[width=.8\linewidth]{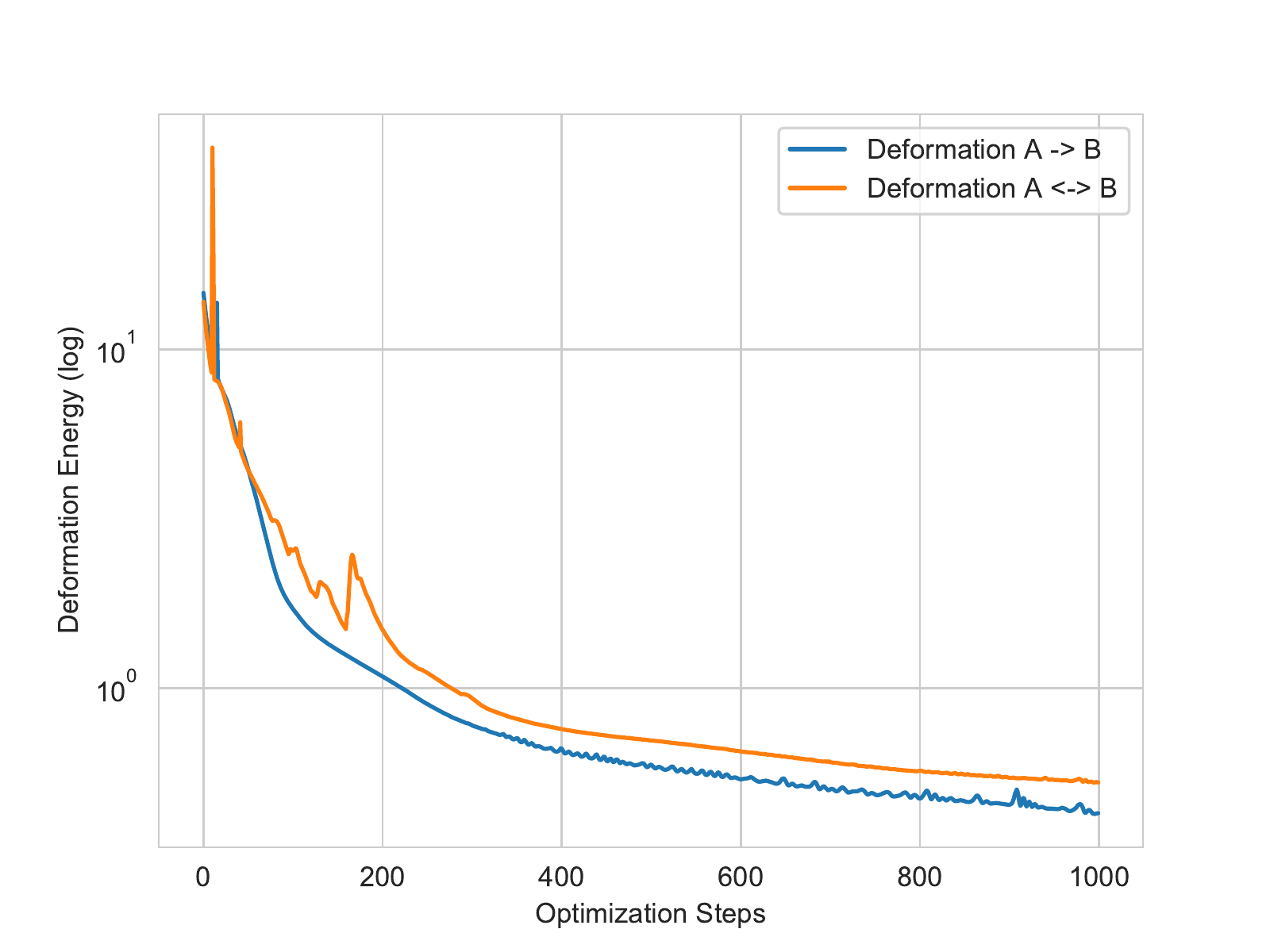}
\caption{We compare two optimizations using Equation~\ref{eq:energy} and~\ref{eq:energy2}, optimizing for single-directional and bi-directional fitting. We evaluate the fitting loss from deformed shape A to B by plotting Equation~\ref{eq:energy}. While optimization with Equation~\ref{eq:energy} is expected to achieve better fitting quality, optimization with Equation~\ref{eq:energy2} achieves similar performance.}
\label{fig:optim}
\end{figure}

\paragraph*{Bidirectional Fitting} Optimization of the single-direction energy (Equation~\ref{eq:energy}) targets a deformation that minimizes the distance between deformed shape A to shape B. In certain scenarios, we want to optimize for bidirectional deformation (Equation~\ref{eq:energy2}) so that either shape can be deformed to fit the other. In Figure~\ref{fig:optim}, we evaluate the fitting loss from deformed shape A to B by plotting Equation~\ref{eq:energy} comparing two different optimizations using Equation~\ref{eq:energy} and~\ref{eq:energy2}. It is expected that optimization with Equation~\ref{eq:energy} achieves lower losses since the objective function is exactly the evaluation metric. However, slightly surprisingly, optimization with Equation~\ref{eq:energy2} achieves a similar fitting quality, suggesting that bidirectional fitting optimization will not affect the fitting quality from either direction. This is due to the intrinsic bijection property of our deformation function.
\section{Application}
\label{sec:application}
In this section, we demonstrate several applications that can benefit from our framework, including novel shape creation, animation, scan-to-CAD fitting, and texture reconstruction. For the shape animation and design, our ODE-based bijective function $\mathcal{D}_{\mathrm{ode}}$ creates bidirectional deformations with smooth intermediate steps (Equation~\ref{eq:energy2}). For scan-to-CAD fitting and texture reconstruction, we choose to optimize for Equation~\ref{eq:energy}. More results can be found in the supplemental material.

\subsection{Shape Animation and Design}
\label{sec:animation}
\begin{figure}
    \centering
    \includegraphics[width=\linewidth]{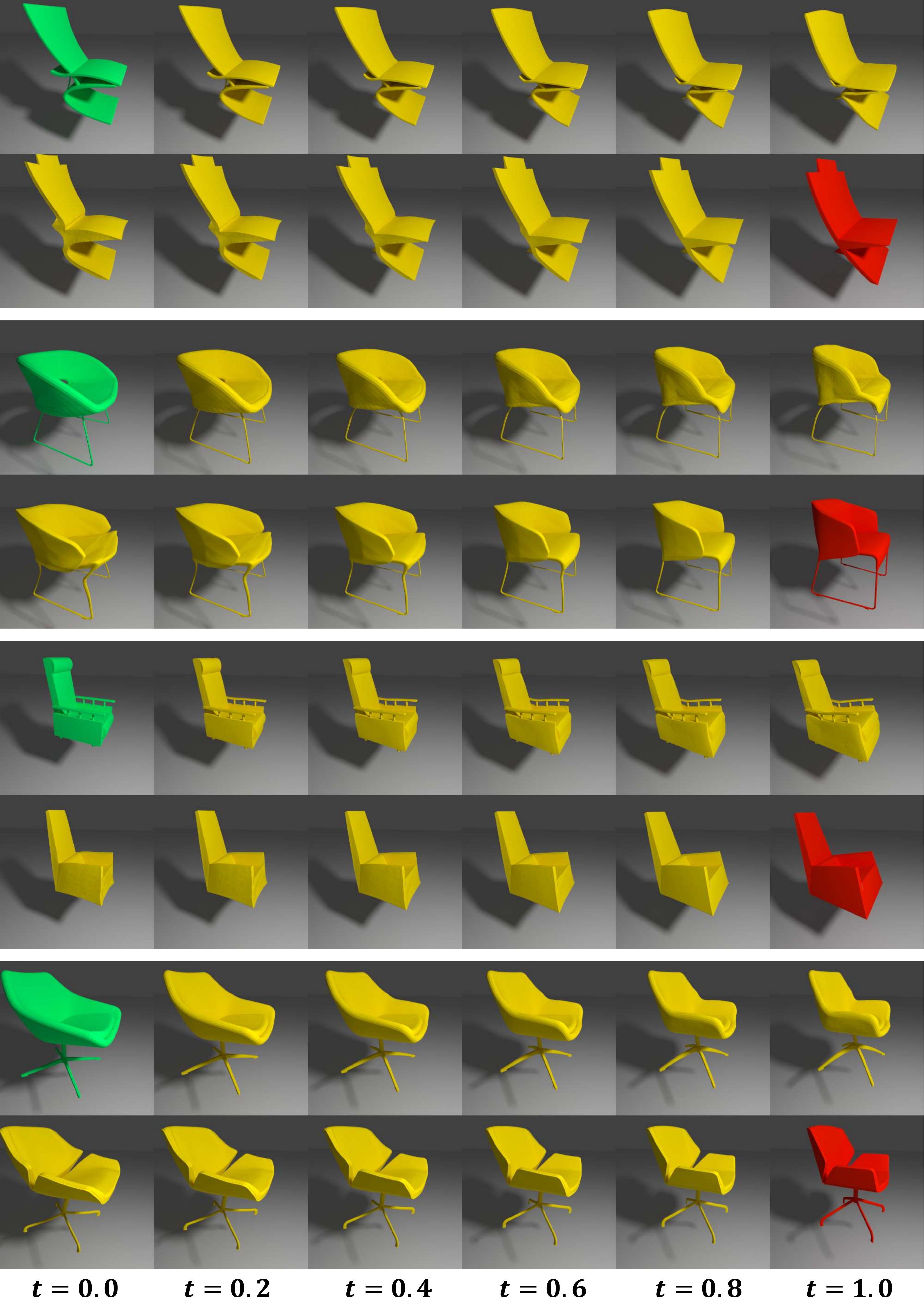}
    \caption{Bidirectional shape animation. For every two rows, our input is the two shapes shown in green and red. Each shape can be deformed to the other with reasonable intermediate steps. Designers can interactively create novel shapes by controlling $t$ with our bijective smooth function.}
    \label{fig:shape-design}
\end{figure}
By optimizing the smooth deformation function $\mathcal{D}_{\mathrm{ode}}$, we are able to obtain a time-dependent velocity fields $\mathbf{u}(\mathbf{p},t)$. We can create intermediate steps of the deformation $\mathbf{p}(\mathbf{x},t_i)$ as an animation, where $\mathbf{x}\in \mathcal{V}_{\mathrm{src}}$ and $t_i$ are uniformly and densely sampled in $[0,1]$. This additionally allows the designer to interactively create new shapes by deciding how close the shape is to the target using the style of the source shape by specifying a single interpolation scalar $t$. Examples of such a novel shape creation process are shown in figure~\ref{fig:shape-design}, where for every two rows our input is the two shapes shown in green and red, and yellow shapes are intermediate results via deformation.

\subsection{Scan-to-CAD}
\label{sec:scantocad}
\begin{figure}
    \centering
    \includegraphics[width=\linewidth]{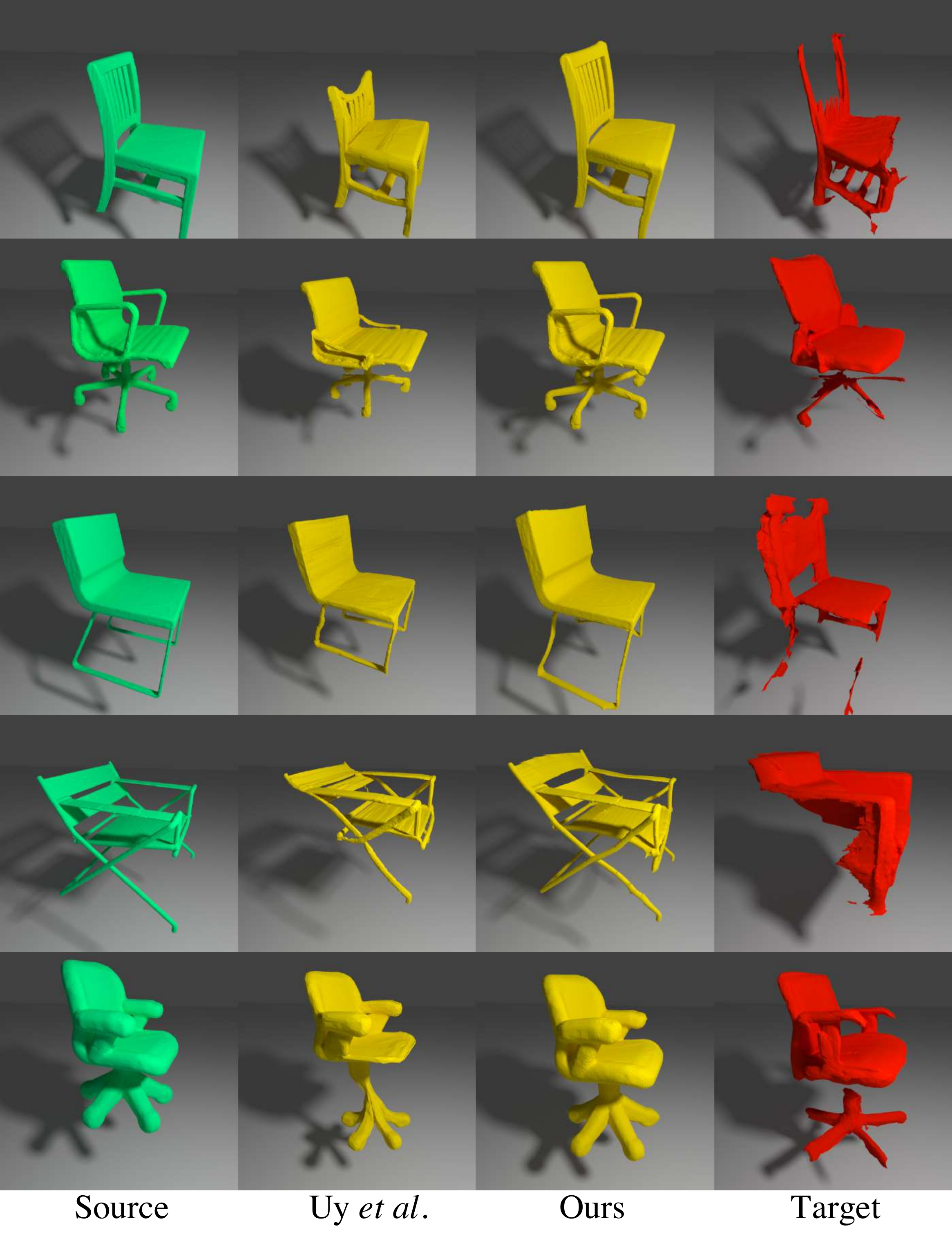}
    \caption{Fitting errors between retrieved/deformed CAD models and scans in Scan2CAD~\cite{avetisyan2019scan2cad}. Our deformation produces results that are more visually appealing with smaller fitting errors, compared to the state-of-the-art.}
    \label{fig:scan2cad-vis}
\end{figure}
\begin{table}
    \centering
    \begin{tabular}{|c|c|c|c|c|}
        \hline
         & Chair & Table & Sofa & All \\
         \hline
         \hline
         Human & 1.715 & 2.832 & 1.947 & 2.119 \\
         \hline
         AE & 1.856 & 2.782 & 2.142 & 2.200 \\
         \hline
         Uy \textit{et al.} & 1.706 & 2.813 & 1.570 & 2.069\\
         \hline
         \hline
         $\mathcal{D}_{\mathrm{ode}}$(Human) & 1.343 & 1.177 & 0.845 & 1.236 \\
         \hline
         $\mathcal{D}_{\mathrm{ode}}$(AE) & 1.401 & 1.104 & 1.051 & 1.264\\
         \hline
         $\mathcal{D}_{\mathrm{ode}}$(Uy \textit{et al.}) & 1.402 & 1.173 & 1.002 & 1.283\\
         \hline
    \end{tabular}
    \caption{we report the fitting error of the deformed CAD model to the scan using different retrieval methods proposed in \cite{uy2020deformation} including human retrieval (Human), nearest neighbor under autoencoder (AE), and their own proposed method (Uy \textit{et al.}). The first three rows use the deformation function in \cite{uy2020deformation} as baselines, and the last three are deformed using our method.
    }
    \label{tab:scan2cad-fit}
\end{table}
By setting the scanned mesh as the target and the CAD model as the source, we can deform the CAD to fit the scan. This is a promising direction to fix the scanning geometry as explored by~\cite{avetisyan2019scan2cad,avetisyan2019end,dahnert2019joint,uy2020deformation}. Using our deformation that minimizes Equation~\ref{eq:energy}, the CAD model can better fit the scanning geometry. In Table~\ref{tab:scan2cad-fit}, we report the fitting error between the deformed CAD model and the scan using different retrieval methods proposed in \cite{uy2020deformation} including human retrieval (Human), nearest neighbor under autoencoder (AE), the retrieval method in Uy \textit{et al.}. The first three rows are use the deformation function in \cite{uy2020deformation} as a baseline, and the last three are deformed using our method. As a result, our deformation produces better results with respect to the fitting error based on different retrieval methods. Figure~\ref{fig:scan2cad-vis} shows some visual comparisons, where we produce deformed source shapes with better fitting quality to scans.

\subsection{Texture Optimization}
Finally, we demonstrate that our deformation framework can help with transferring scanned textures to scanned CAD models that are aligned with the scans. This can be used for enhancing the appearance of the CAD fitting to the scans, as proposed in~\cite{huang2020adversarial}.

\begin{figure}
    \centering
    \includegraphics[width=\linewidth]{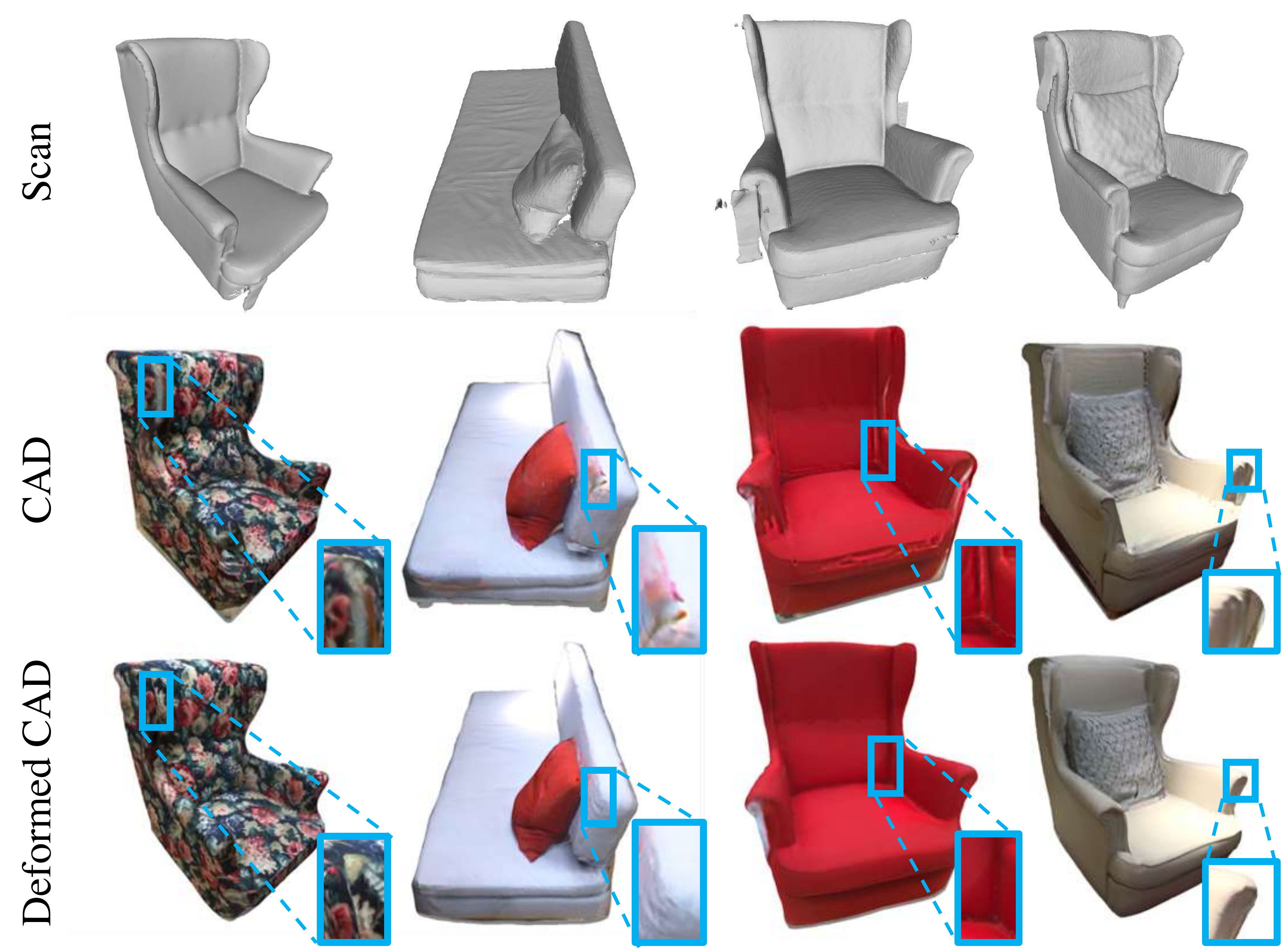}
    \caption{Texture optimization comparisoned with~\cite{huang2020adversarial}. Texture can be better optimized with our deformed CAD model since it fits better to the scan and has better geometry alignment with the images for the scan.}
    \label{fig:texture-optim}
\end{figure}
With good deformation, the texture is easier to optimize from observed images to the fitted CAD model, since our deformation enables better alignment and therefore less geometry alignment errors. Specifically, we optimize based on the two-way fitting (Equation~\ref{eq:energy}) to deform the CAD to the scanned geometry and optimize the texture from images to the geometry according to to~\cite{huang2020adversarial}. Compared to the result from~\cite{huang2020adversarial} without any deformation, the texture appearance on our deformed CAD model has fewer artifacts, as shown in Figure~\ref{fig:texture-optim}. Such improvement mostly happen at the corners or object boundaries, where severely misaligned regions cannot be easily address by pure texture optimization.

\section{Conclusion}
We develop a robust and scalable framework for CAD deformation. Our key contribution is a novel neural deformation function that enables bi-directional deformation between two shapes with continuous intermediate steps. Our mesh representation enables valid deformation with complex CAD model structures and our results significantly outperform existing state-of-the-art methods. Our deformation framework enables several applications related to shape animation and design, scanning, and texture reconstruction.

Our method can be further extended in several future directions. A multi-grid optimization could further improve the deformation in a global-local manner with better convergence. It is possible to enforce a divergence-free velocity field by applying a curl operator. This will be essential for applications that require volume-preserving deformation, e.g., human animation. Finally, it is possible to modify the velocity field of the neural function to accept additional parameters to describe deformations between multiple models and encode the models into a deformation space. 

\bibliographystyle{ACM-Reference-Format}
\bibliography{sample-base}

\clearpage
\section*{Appendix}
\begin{appendix}

\section{Shape Deformation}
\begin{figure*}
\includegraphics[width=\linewidth]{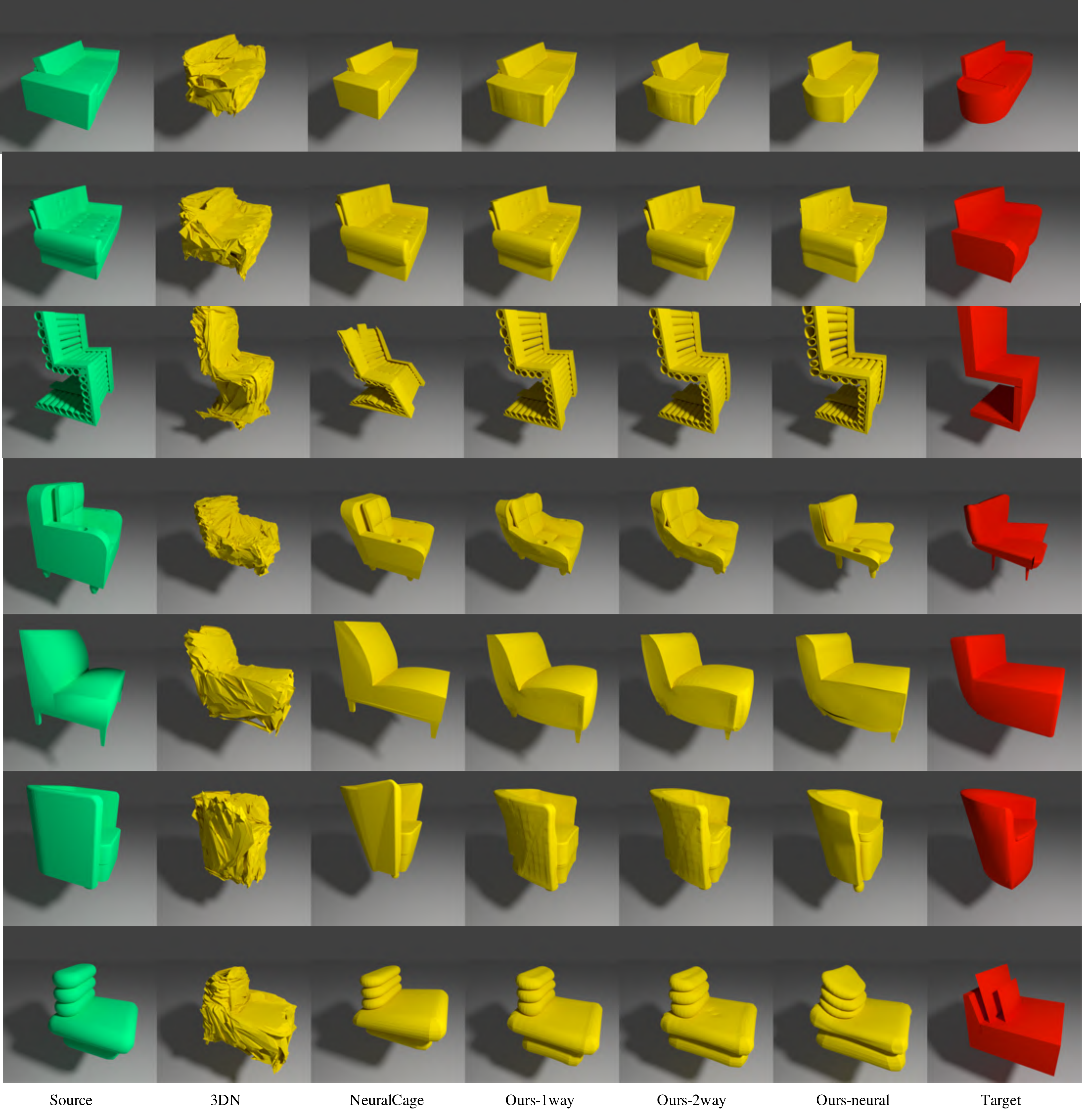}
\caption{Visual comparison on shape deformation between different approaches (examples set 2).}
\label{fig:supp-deform1}
\end{figure*}
\begin{figure*}
\includegraphics[width=\linewidth]{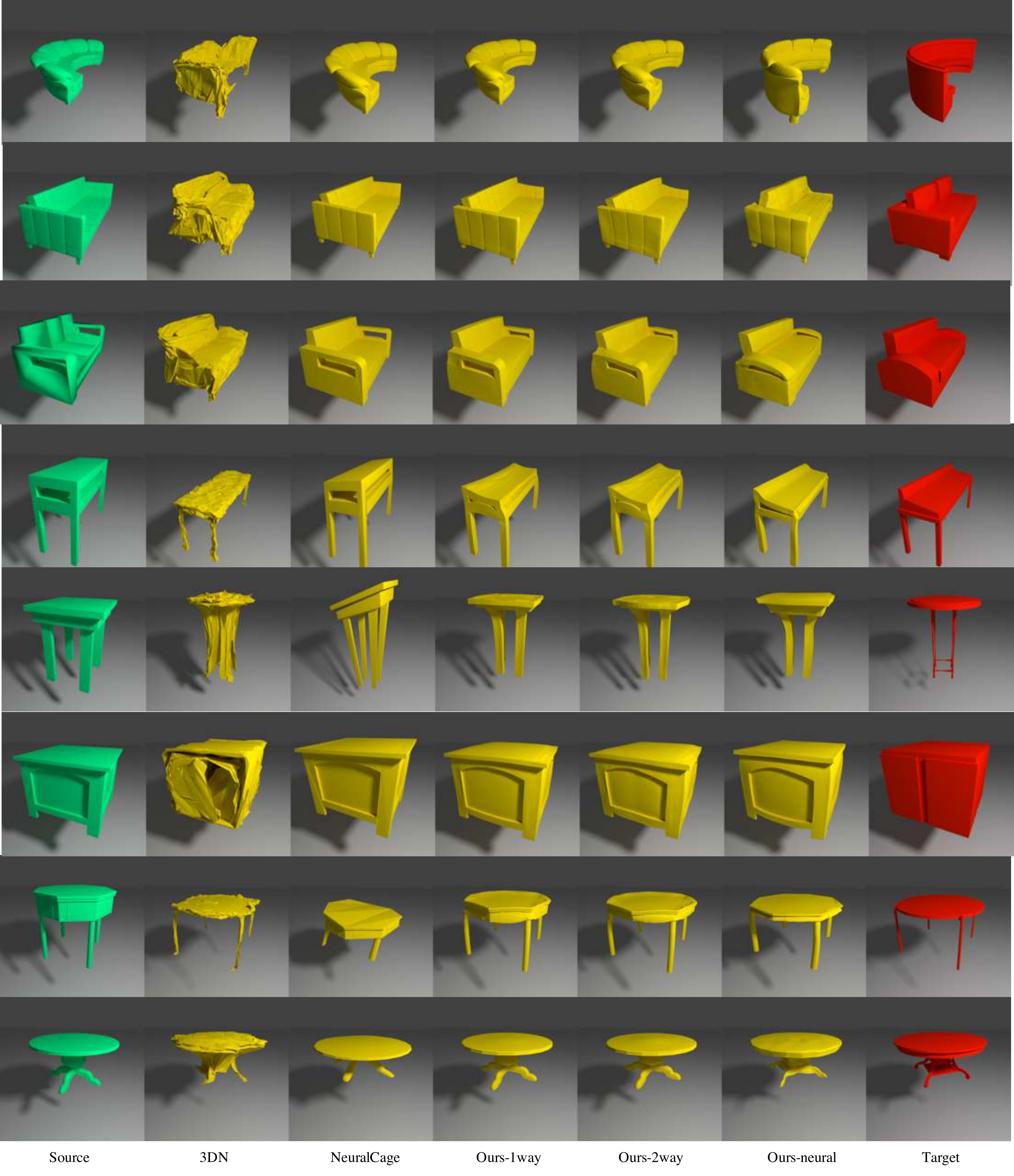}
\caption{Visual comparison on shape deformation between different approaches (examples set 2).}
\label{fig:supp-deform2}
\end{figure*}
Following Section 6.1 in the main paper, we provide additional visual comparisons on shape deformation in Figure~\ref{fig:supp-deform1} and~\ref{fig:supp-deform2}. 3DN~\cite{wang20193dn} and NeuralCage~\cite{yifan2019neural} do not fit the target as good as our methods do. Ours-1way is not optimized for the coverage of the target, while ours-2way ensures coverage of both the deformed and the target shape. Ours-neural additionally provides more flexible deformation with better fitting quality.

\section{Shape Creation and Animation}
\begin{figure*}
\includegraphics[width=0.79\linewidth]{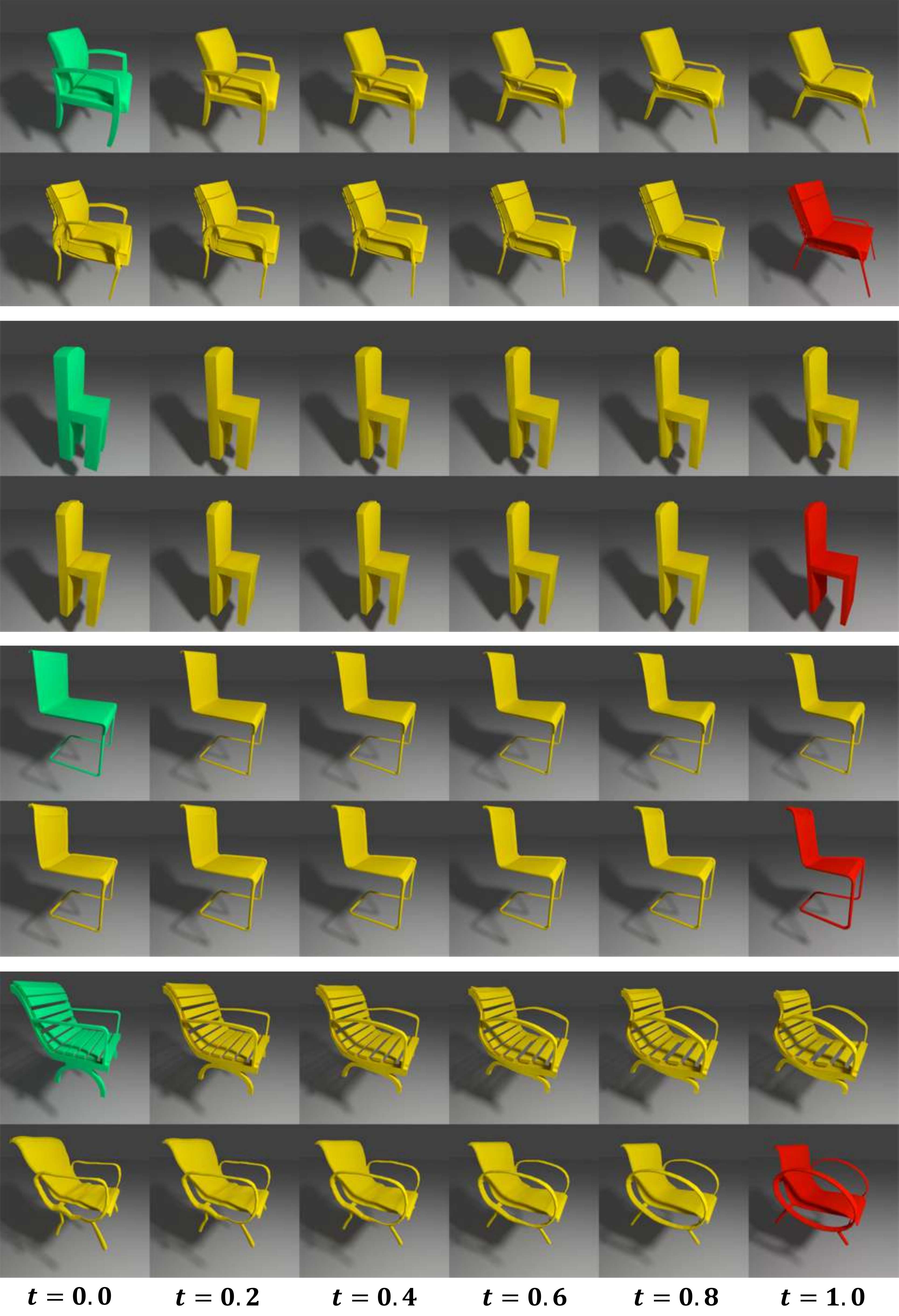}
\caption{Bidirectional shape animation (examples set 1). For every two rows, our input is the two shapes shown in green and red. Each shape can be deformed to the other with reasonable intermediate steps. Designers can interactively create novel shapes by controlling $t$ with our bijective smooth function.}
\label{fig:supp-animate1}
\end{figure*}
\begin{figure*}
\includegraphics[width=0.8\linewidth]{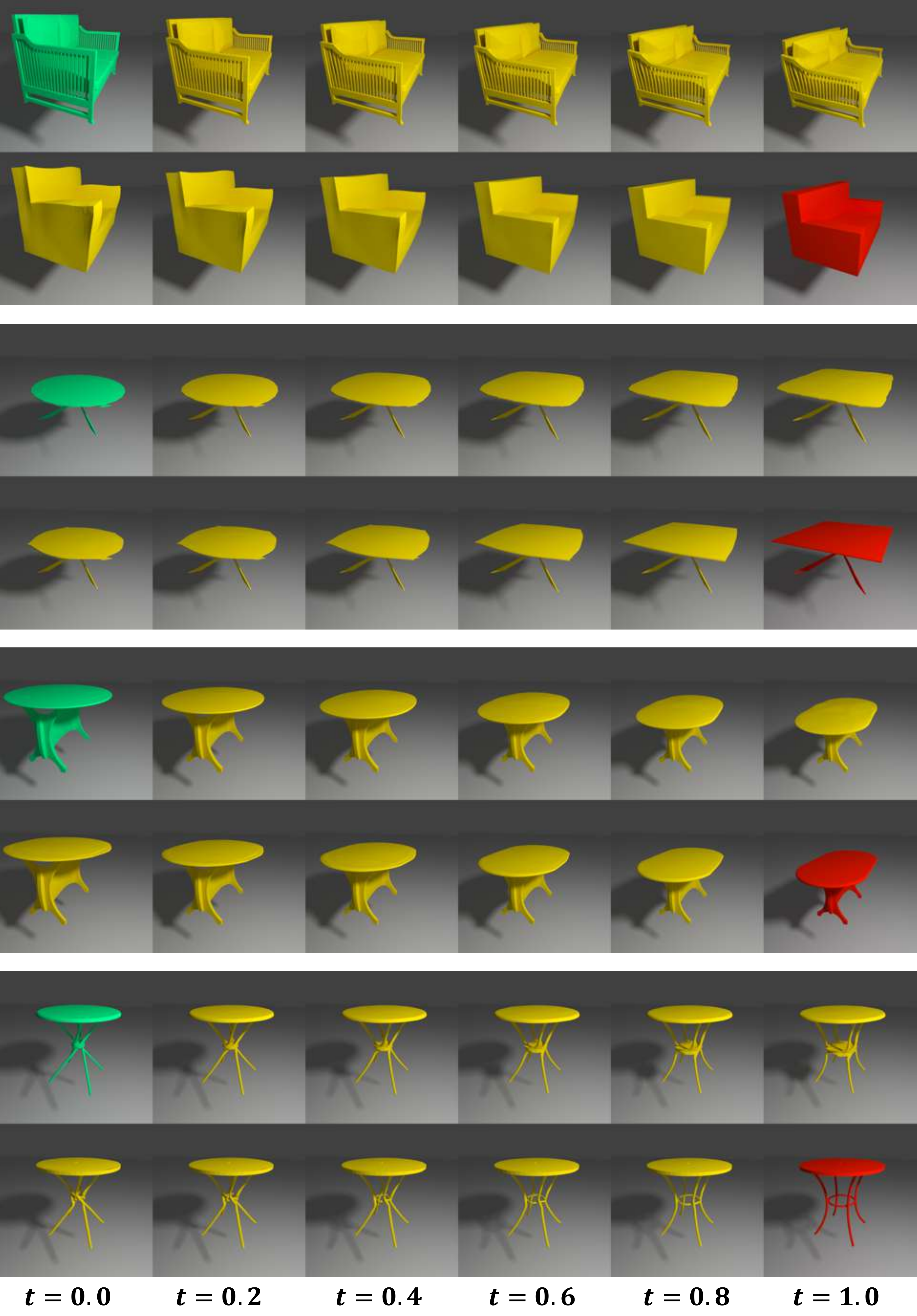}
\caption{Bidirectional shape animation (examples set 2). For every two rows, our input is the two shapes shown in green and red. Each shape can be deformed to the other with reasonable intermediate steps. Designers can interactively create novel shapes by controlling $t$ with our bijective smooth function.}
\label{fig:supp-animate2}
\end{figure*}
\begin{figure*}
\includegraphics[width=0.8\linewidth]{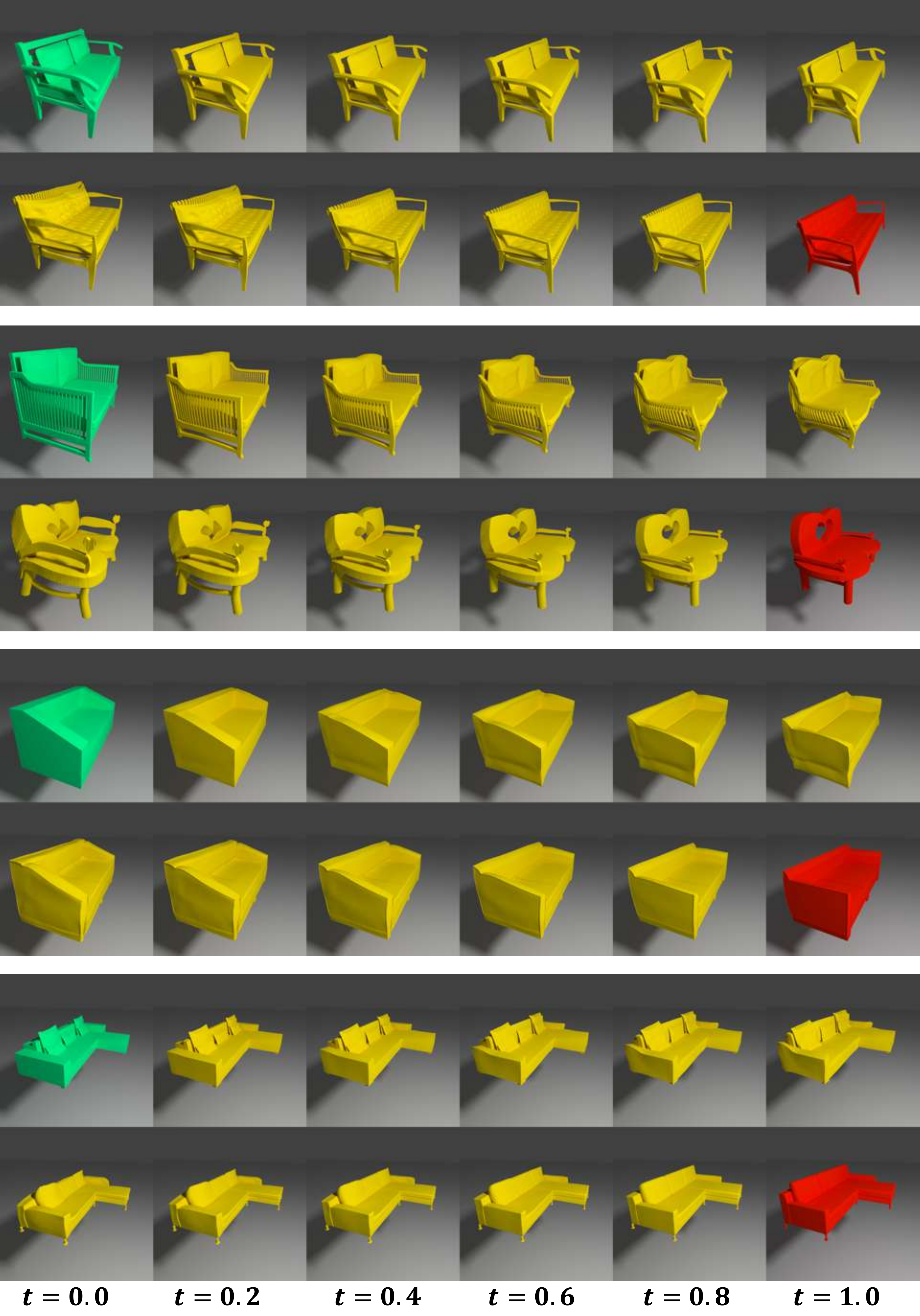}
\caption{Bidirectional shape animation (examples set 3). For every two rows, our input is the two shapes shown in green and red. Each shape can be deformed to the other with reasonable intermediate steps. Designers can interactively create novel shapes by controlling $t$ with our bijective smooth function.}
\label{fig:supp-animate3}
\end{figure*}
Following Section 7.1 in the main paper, we provide additional examples for shape animation and design in Figure~\ref{fig:supp-animate1},~\ref{fig:supp-animate2} and~\ref{fig:supp-animate3}. By optimizing the smooth deformation function $\mathcal{D}_{\mathrm{ode}}$, we are able to obtain a time-dependent velocity field $\mathbf{u}(\mathbf{p},t)$. We can create intermediate steps of the deformation $\mathbf{p}(\mathbf{x},t_i)$ as an animation, where $\mathbf{x}\in \mathcal{V}_{\mathrm{src}}$ and $t_i$ are uniformly and densely sampled in $[0,1]$. Such intermediate steps can be used for interactive novel shape design.

\section{Scan-to-CAD Fitting}
\begin{figure*}
\includegraphics[width=0.8\linewidth]{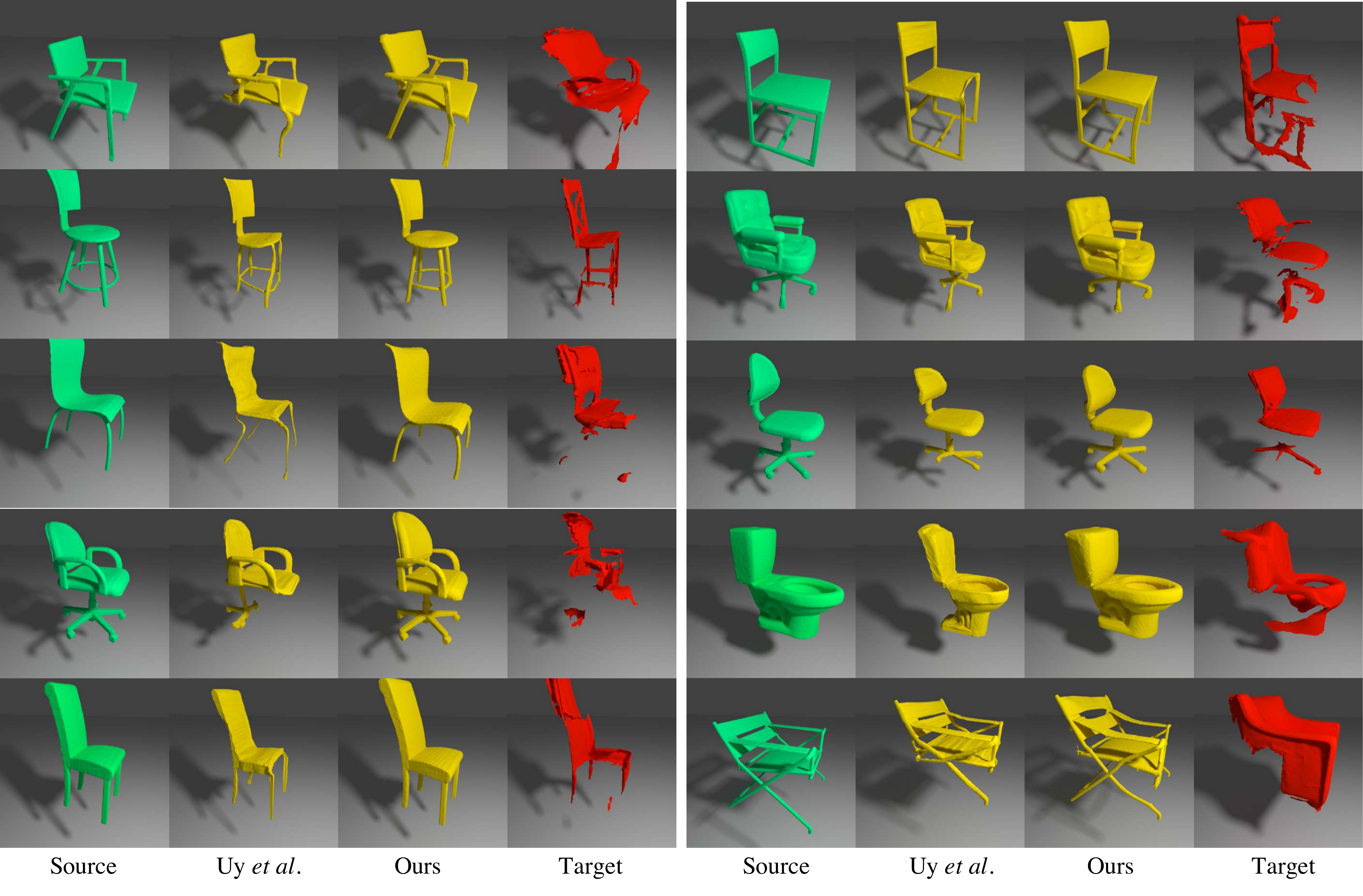}
\caption{Fitting errors between retrieved/deformed CAD models to scans in Scan2CAD~\cite{avetisyan2019scan2cad}. Our deformation produces less fitting errors and more visually appealing results compared to the state-of-the-art.}
\label{fig:supp-scan2cad}
\end{figure*}
Following Section 7.2 in the main paper, we provide additional examples for scan-to-CAD deformation fitting comparisons in Figure~\ref{fig:supp-scan2cad}. Comparing with~\cite{uy2020deformation}, we produce better fitting results without collapsing parts which are missing in scans.

\section{Texture Reconstruction}
\begin{figure*}
\includegraphics[width=0.7\linewidth]{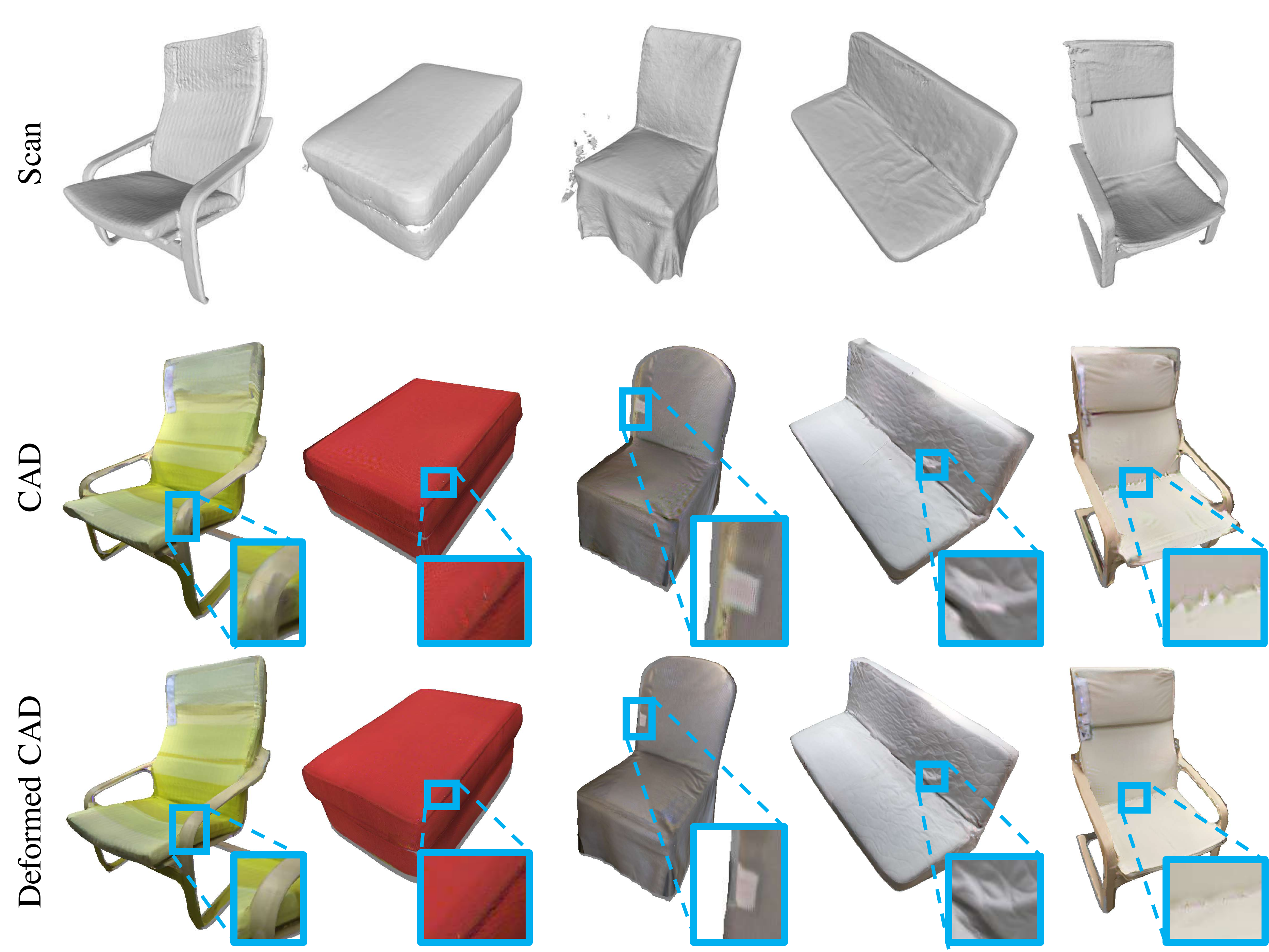}
\caption{Texture optimization comparison with~\cite{huang2020adversarial}. Texture can be better optimized with our deformed CAD model since it fits better to the scan and has better geometry alignment with the scanning images.}
\label{fig:supp-texture}
\end{figure*}
Following Section 7.3 in the main paper, we provide additional examples in Figure~\ref{fig:supp-texture} comparing texture reconstruction using the roughly-aligned original and deformed CAD models using~\cite{huang2020adversarial}. Deformed CAD models fit scans better and leads to better texture reconstruction from aligned images.

\end{appendix}

\end{document}